\documentclass[prb,twocolumn,showpacs,preprintnumbers,superscriptaddress]{revtex4}
\usepackage{graphicx}
\usepackage{dcolumn}
\usepackage{subfigure}
\usepackage{wrapfig}
\usepackage{cancel}
\usepackage{color}
\usepackage{textcomp}
\usepackage{amsmath}
\usepackage{amsfonts}
\usepackage{amssymb}

\newcommand{\nio}{Na$_2$IrO$_3$}
\newcommand{\eg}{\it{e.~g.}}
\newcommand{\ie}{\it{i.~e.}}

\begin{document}

\title{\textit{Ab initio} analysis of the tight-binding parameters and
magnetic interactions in {Na$_{2}$IrO$_{3}$}}
\author{Kateryna Foyevtsova}
\affiliation{Institut f\"ur Theoretische Physik, Goethe-Universit\"at Frankfurt, 60438
Frankfurt am Main, Germany}
\author{Harald O. Jeschke}
\affiliation{Institut f\"ur Theoretische Physik, Goethe-Universit\"at Frankfurt, 60438
Frankfurt am Main, Germany}
\author{I. I. Mazin}
\affiliation{Code 6393, Naval Research Laboratory, Washington, DC 20375, USA}
\author{D. I. Khomskii}
\affiliation{II. Physikalisches Institut, Universit\"at zu K\"oln, Z\"ulpicher Strasse
77, 50937 K\"oln, Germany}
\author{Roser Valent\'{\i}}
\affiliation{Institut f\"ur Theoretische Physik, Goethe-Universit\"at Frankfurt, 60438
Frankfurt am Main, Germany}

\date{\today }

\begin{abstract}
  By means of density functional theory (DFT) calculations (with and
  without inclusion of spin-orbit (SO) coupling) we present a detailed
  study of the electronic structure and corresponding microscopic
  Hamiltonian parameters of {\nio}.  In particular, we address the
  following aspects: (i) We investigate the role of the various
  structural distortions and show that the electronic structure of
  {\nio} is exceptionally sensitive to structural details.  (ii) We
  discuss both limiting descriptions for {\nio}; quasi-molecular
  orbitals (small SO limit, itinerant) versus 
  relativistic orbitals (large SO limit, localized) and show that the
  description of {\nio} lies in an intermediate regime.  (iii) We
  investigate whether the nearest neighbor Kitaev-Heisenberg model is
  sufficient to describe the electronic structure and magnetism in
  {\nio}. In particular, we verify the recent suggestion of an
  antiferromagnetic Kitaev interaction and show that it is not
  consistent with actual or even plausible electronic
  parameters. Finally, (iv) we discuss correlation effects in
  {\nio}. We conclude that while the Kitaev-Heisenberg Hamiltonian is
  the most general expression of the quadratic spin-spin interaction
  in the presence of spin-orbit coupling (neglecting single-site
  anisotropy), the itinerant character of the electrons in {\nio}
  makes other terms beyond this model (including, but not limited to
  2nd and 3rd neighbor interactions) essential.
\end{abstract}

\maketitle

\section{Introduction}

The electronic and magnetic behavior of layered $5d$ transition metal
oxides~\cite{Kim_2009} has been a subject of intensive discussion in
the last years. Particularly exciting has been the suggestion by the
authors of Ref.~\onlinecite{Chaloupka2010} that hexagonal iridates
such as {\nio} are a realization of the nearest neighbor
Kitaev-Heisenberg (nnKH) model:
\begin{equation}
H_{ij}^{(\gamma )}=2KS_{i}^{\gamma }S_{j}^{\gamma }+J{\bf S}_{i}\cdot 
{\bf S}_{j}  \label{nnKHmodel}
\end{equation}
This proposal is based on the premise that spin-orbit (SO) coupling is
the most important energy scale for the description of these systems
so that Ir $5d$ $t_{2g}$ orbitals are written in terms of
$j_{\text{eff}}=1/2$ and $j_{\text{eff}}=3/2$ relativistic orbitals,
with the Kramers doublet $j_{\text{eff}}=1/2$ represented by the
operator $S=1/2$. The combination of Kitaev and Heisenberg terms leads
to a complex phase diagram with various magnetic and spin-liquid
phases~\cite{Chaloupka2010,trebst1,Chaloupka2012}. Obviously, some of
these properties can only manifest themselves when the Kitaev term
dominates or is at least comparable to the Heisenberg term. Also,
other possible contributions, such as magnetic anisotropy, ring
exchange or biquadratic exchange, to mention a few, may alter the
phase diagram and the properties of the model considerably. Most
importantly, while the Kitaev-Heisenberg expression is the most
general fully-symmetric expression for anisotropic pairwise magnetic
interactions in the second order in spin in the presence of SO coupling
(just as the Heisenberg exchange represents the same in the isotropic
non-relativistic case), it is not necessarily short ranged in the
presence of considerable itinerancy.

So far, essentially all analyses of the nnKH model for {\nio} have
been performed in the localized limit, where an assembly of weakly
interacting relativistic atomic orbitals is assumed to be a good
starting approximation. On the other hand, first principles
calculations suggest considerable delocalization of electrons over
individual Ir hexagons building quasi-molecular orbitals
(QMOs)~\cite{we}. The associated
{\textquotedblleft}itinerant{\textquotedblright} energy scale (the
band width) is $\approx1.5$~eV, to be compared to the single-site
spin-orbit splitting scale~\cite{comment_split}
 $(3/2)\lambda\approx0.7$~eV and the Hubbard
and Hund's rule correlation energy scale of
$U-J_{\mathrm{H}}\approx0.5-1$~eV. This makes the entire premise of
the nnKH model questionable. At the same time, it has also been
pointed out~\cite{trebst,Radu} that the nnKH model with the addition
of the 2nd and 3rd neighbors Heisenberg interaction is easier to
reconcile with the experimental data. Such relatively long-range exchange
interaction is another hallmark of considerable itinerancy (here and
below, when we speak of itinerancy, we imply mostly delocalization
over Ir$_6$ rings, but not necessarily over the entire crystal).

In the present work we revisit and discuss the validity of both
limiting descriptions for {\nio}; itinerant (QMO picture) versus
localized ($j_{\text{eff}}=1/2$ Kramers doublet). To this end, we
perform a thorough analysis of the electronic properties of {\nio}
within {\it non-relativistic} and {\it relativistic}
 density functional theory
(DFT) and derive, using projection on Wannier functions, the relevant
hopping parameters and show that QMOs are naturally obtained as linear
combinations of Ir $t_{2g}$ Wannier functions. We discuss the relation
between the quasi-molecular orbital and the relativistic orbital,
 $j_{\text{eff}}$, representations and show that
the behavior of {\nio} lies in between a fully localized and fully
itinerant description.  Finally, the parametrization of the electronic
bandstructure allows us to provide realistic estimates for the model
parameters in the localized nnKH model. We thus investigate whether we
are close to a regime where the Kitaev interaction plays a decisive
role or not.

Quite unexpectedly, we find that {\nio} is an example of a material
where minor details of the crystal structure can dramatically affect
the electronic structure, and simple guessing of the band structure
parameters, or estimating them from simplified crystallographic models
(so far all model calculations for this compound were utilizing one or
the other approach) can be exceptionally misleading. In fact some of
the models energetically discussed in the community, while of
undeniable theoretical appeal, are not even qualitatively close to the
actual parameter range in {\nio}.

While this particular compound is very intriguing and has been
enjoying extraordinary popularity lately, we want to emphasize that
this strong dependence of the electronic properties on details of the
crystal structure is an important result, whose relevance goes beyond
specifically {\nio} and is likely true for many other materials based
on honeycomb transition-metal layers.

The paper is organized as follows. In Section~\ref{sec:two} we review
the crystal structure and magnetic properties of {\nio}. In
Section~\ref{sec:three} we provide details of the DFT calculations and
the projector method. In Section~\ref{sec:four} we present the results
of the electronic structure analysis without inclusion of spin-orbit
coupling and analyze the role of the structural distortions in
{\nio}. In Section~\ref{sec:five} we investigate the role of
spin-orbit coupling and discuss the relation between the QMOs and the
relativistic orbitals ($j_{\text{eff}}$). In this context, we discuss
whether the existing experimental situation can distinguish between
the DFT description (with the resulting itinerancy) and localized
($j_{\mathrm{eff}}=1/2)$ models. We proceed with an analysis of the
single-site magnetic anisotropy in {\nio} and find it to be relevant
(pure $j_{\mathrm{eff}}=1/2$ states do not have any single-site
anisotropy).
In Section~\ref{sec:six} we provide \textit{ab initio}-derived
estimates for the parameters appearing in the Kitaev and Heisenberg
terms in {\nio} and discuss the validity of the nnKH model by
considering the experimentally observed magnetic order and attempts to
explain it from a local point of view. Finally in
Section~\ref{sec:seven} we present our conclusions.
\begin{figure}[ptb]
\begin{center}
\includegraphics[width=0.95\columnwidth]{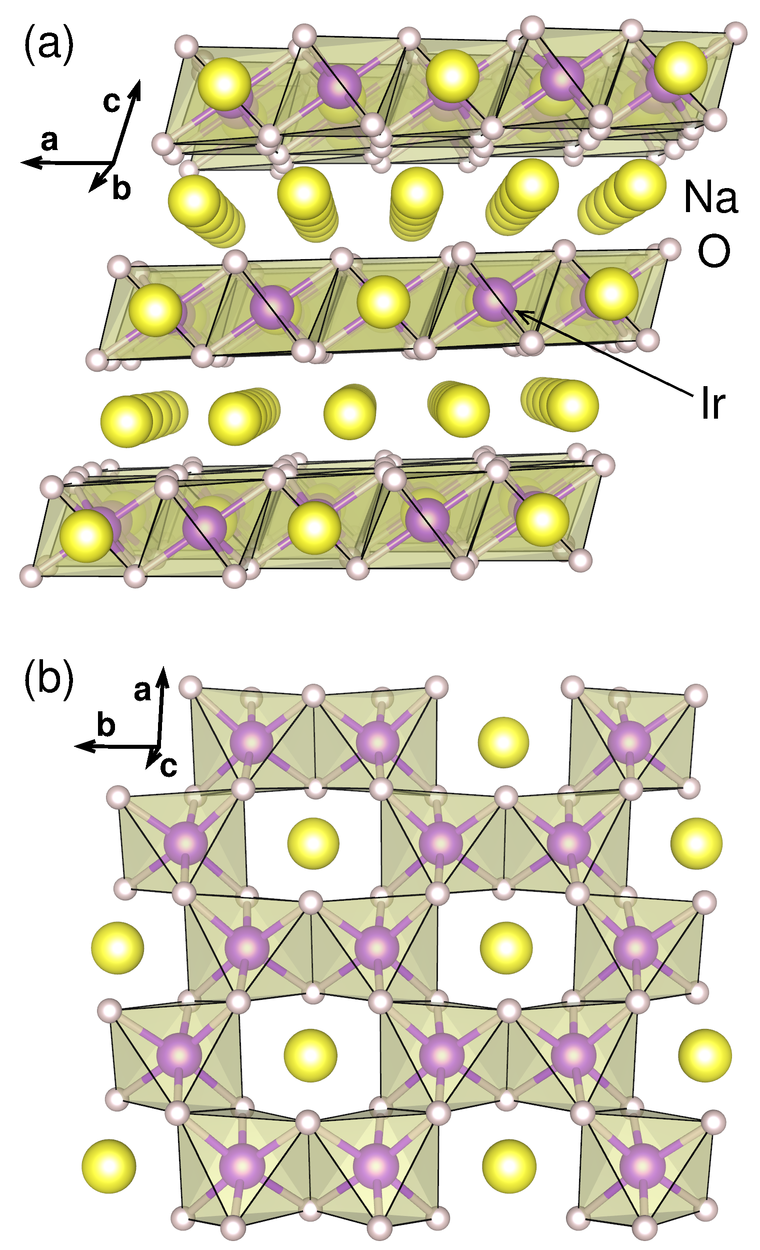}
\end{center}
\caption{ Crystal structure of {\nio}.  (a) Projection on the $ac$
  plane and (b) projection on the $ab$ plane. }
\label{crystal}
\end{figure}

\section{Crystal structure and magnetic properties of
  {N\lowercase{a}$_2$I\lowercase{r}O$_3$}}
\label{sec:two}

{\nio} crystallizes in the monoclinic space group $C\,2/m$
(No.~12)~\cite{Radu} (see Fig.~\ref{crystal}) and consists of Ir
honeycomb layers (Fig.~\ref{crystal} (b)) stacked along the monoclinic
{\bf c} axis (Fig.~\ref{crystal} (a)) with an in-plane off-set along
{\bf a}. Na ions occupy both the interlayer positions and 1/3 of the
in-plane positions at the centers of Ir hexagons. This structure can
be visualized as proceeding from NaIrO$_{2}$ with a CdI$_{2}$
structure with triangular IrO$_{2}$ layers. In these layers 1/3 of the
in-plane iridium atoms are substituted by extra Na, {\ie}, its formula can
be written as Na(Na$_{1/3}$Ir$_{2/3})$O$_{2}$, which, multiplied by
3/2, gives the usual formula of {Na$_{2}$IrO$_{3}
  $}~\cite{comment_cava}.

An idealized crystal structure of this kind corresponds to having all
nearest neighbor (NN) Ir-Ir and NN Ir-O distances equal and Ir-O-Ir
angles of 90 degrees. The experimental structure of {\nio} departs
from the idealized case and shows a few distortions: (i) orthorhombic
distortion that introduces inequality among NN Ir-Ir distances and
among NN Ir-O distances, (ii) IrO$_{6}$ octahedra rotations that place
O atoms on the faces of a cube containing an Ir hexagon (see Fig. 2 of
Ref.~\onlinecite{we}) and (iii) trigonal distortion which is a
compression of the IrO$_{6}$ octahedra in the {\bf c}-direction that
induces a departure from 90 degrees of the Ir-O-Ir angles. In
Section~\ref{sec:four} we will discuss the effect of these
distortions on the electronic structure of {\nio}.

As shown by transport, optical and high-energy spectroscopy
studies~\cite{Singh10,Comin12}, {\nio} is an insulator with an energy
gap $E_{g}$ of 340~meV. Magnetic susceptibility measurements indicate
a Curie-Weiss behavior at high temperatures with a Curie-Weiss
temperature $\Theta _{CW}=-116$~K and an effective Ir moment
$\mu_{\text{eff}}=1.82\mu_{{\text{B}}}$. {\nio} orders
antiferromagnetically below $T_{N}=15$~K with an ordered magnetic
moment $\mu_{ord} \sim 0.2\mu_{\text{B}}$. The fact that $T_{N}$ is
much smaller than $\Theta _{CW}$ may be a signature of frustration,
but it may be also caused by the itinerancy of Ir $5d$
electrons~\cite{we} as will be discussed in Section~\ref{sec:six}.

\begin{figure}[ptb]
\includegraphics[width=0.5\textwidth]{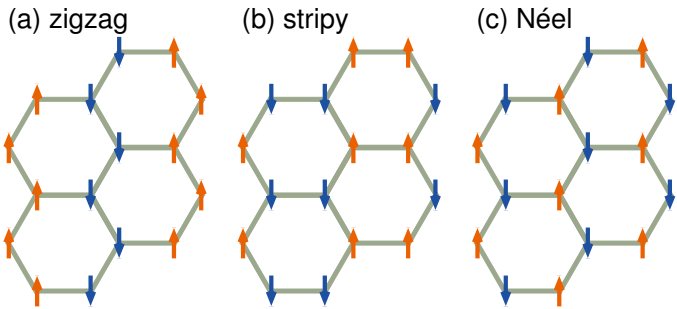}
\caption{Possible antiferromagnetic patterns in a honeycomb lattice}
\label{patterns}
\end{figure}

The magnetic pattern observed experimentally~\cite{Radu} corresponds
to a zigzag ordering, in contrast to the prediction of a stripe order
by the nnKH model~\cite{Chaloupka2010} (see Fig.~\ref{patterns}).
Recently, Chaloupka \textit{et al.}~\cite{Chaloupka2012} argued that
such a zigzag ordering can be also obtained by the nnKH model, when
one correctly includes all the terms contributing to NN Ir-Ir
exchange. In Section~\ref{sec:five}, we will discuss this proposition
in more detail.

\section{Method}
\label{sec:three}

 In this work we perform DFT calculations
using the linearized augmented plane wave (LAPW) method as implemented
in the full-potential code WIEN2k~\cite{w2k}. We employ the 
Perdew-Burke-Ernzerhof
generalized gradient approximation~\cite{PBE} to the DFT
exchange-correlation functional and set the basis-size controlling
parameter $RK_{\text{max}}$~\cite{Cottenier} to 7. We consider a mesh
of 500 ${\bf k}$-points in the first Brillouin zone. Relativistic
effects are treated within the second variational
approach. Convergence with respect to relevant parameters (the
${\bf k}$-point mesh, the $RK_{\max }$ and the second variational
energy cutoff, \textit{etc.}) has been carefully checked.

\subsection{Calculation of hopping integrals}

In order to be able to discuss various Ir-Ir $5d$ processes, we
parameterize our non-relativistic
DFT results in terms of a tight-binding (TB) model
where the TB Ir $5d$ hopping parameters are obtained through the
Wannier function projection formalism proposed in
Ref.~\onlinecite{Aichhorn09} and generalized to molecular Wannier functions
in Ref.~\onlinecite{Ferber12}.  We first construct Wannier function
projectors $P^{\alpha}_{m\nu}({\bf k})$ for the three $t_{2g}$ Ir~$5d$
orbitals and calculate the TB Hamiltonian $H^{\text{TB}}({\bf k})$ (in
matrix form) via
\begin{equation}
H^{\text{TB}}({\bf k}) = P({\bf k}) D({\bf k}) P^{\dagger}({\bf k}),
\end{equation}
where $D({\bf k})$ is a diagonal matrix of Ir~$5d$ $t_{2g}$ Bloch
eigenvalues and the matrix $P({\bf k})$ is formed by the projectors
$P^{\alpha}_{m\nu}({\bf k}) $. Here, indices $\alpha$, $m$, and $\nu$
run over equivalent Ir atoms in the unit cell,
Ir~$t_{2g}$ orbitals, and Bloch bands, respectively. {\nio} has two Ir per unit cell
and only the six Ir~$t_{2g}$ bands near the Fermi level $E_{\text{F}}$
are considered in the construction of projectors.

We calculate the hopping integral $t^{mm^{\prime}}_{\alpha- {\bf R},
  \alpha^{\prime}- {\bf R}^{\prime}}$ between orbital $m$ on Ir atom
$\alpha$ in the unit cell at a distance ${\bf R}$ from a reference
unit cell and orbital $m^{\prime}$ on Ir atom $\alpha^{\prime}$ in the
unit cell at a distance ${\bf R}^{\prime}$ from a reference unit cell
by integrating $H^{\text{TB}}({\bf k})$ over $N_{{\bf k}}$
${\bf k}$-vectors in the first Brillouin zone:
\begin{equation}
t^{mm^{\prime}}_{\alpha- {\bf R}, \alpha^{\prime}- {\bf R}^{\prime}} = \frac{1}{N_{{\bf k}}} \sum_{{\bf k}} H^{\text{TB}}_{\alpha m, \alpha^{\prime
}m^{\prime}}({\bf k}) e^{-i{\bf k}({\bf R}-{\bf R}^{\prime})}.
\label{TB_param}
\end{equation}
where $H^{\text{TB}}_{\alpha m, \alpha^{\prime}m^{\prime}}({\bf k})$
are the matrix elements of $H^{\text{TB}}({\bf k})$.  Correspondingly,
the diagonal matrix elements $t^{mm}_{\alpha\alpha}$ give the on-site
energies.

\subsection{Construction of quasi-molecular projectors}

As was argued in Ref.~\onlinecite{we}, the most natural description of
the electronic structure of {\nio} is in terms of quasi-molecular (QMO)
orbitals localized on a hexagon. The strongest Ir-Ir hopping is
between $5d$ $t_{2g}$ orbitals of neighboring iridium ions via common
oxygens. In this case,  an electron on a given Ir $t_{2g}$ orbital
propagates around an Ir$_{6}$ hexagon with the peculiarity
than only a certain  $t_{2g}$
orbital at each Ir participates in the hopping~\cite{axes}, e.g.
Ir1(${xy}$)-Ir2(${xz}$)-Ir3(${yz}$)-Ir4(${xy}$)-Ir5(${xz}$)-
Ir6(${yz}$) (see Fig.~2 of Ref.~\onlinecite{we}).  These QMOs are analogous
to the molecular orbitals of the benzene molecule C$_{6}$H$_{6}$
except for the fact that in benzene the same $p$-orbital on each
carbon ion participates in the formation of the molecular orbital
while in {\nio}, as described above, different $t_{2g}$ orbitals are
involved in one QMO and the three $t_{2g}$ orbitals on one Ir ion
contribute to three different neighboring QMOs. We elaborate the
details of the construction of the QMOs in what follows.

QMO projectors $P_{\mathcal{M}\nu }({\bf k})$ are obtained as linear
combinations of Ir~$t_{2g}$ projectors $P_{M\nu }({\bf k})$:
\begin{equation}
P_{\mathcal{M}\nu }({\bf k})=\sum_{M}U_{\mathcal{M},M}T_{M}({\bf k})P_{M\nu
}({\bf k}).\label{E.QMOproj}
\end{equation}
where in the Ir~$t_{2g}$ projectors $P_{M\nu }({\bf k})$, the index
$M$ combines now the atomic index $\alpha $ and orbital index $m$,
{\textit{i.e.}} $M$ runs over all $t_{2g}$ orbitals of all equivalent
Ir atoms. With QMOs ordered as
$\mathcal{M}=A_{1g},E_{2u},E_{1g},B_{1u},E_{1g},E_{2u}$ and
Ir~$t_{2g}$ orbitals ordered as
$M={xy}^{1},{xz}^{1},{yz}^{1},{xy}^{2},{xz}^{2},{yz}^{2}$ (the upper
index labels Ir atoms), $U$ is given by [$\omega =\exp (i\pi /3)$]
\begin{equation}
U=
\begin{pmatrix}
1 & 1 & 1 & 1 & 1 & 1 \\ 
1 & \omega ^{4} & \omega ^{2} & -1 & \omega  & \omega ^{5} \\ 
1 & \omega ^{2} & \omega ^{4} & 1 & \omega ^{2} & \omega ^{4} \\ 
1 & 1 & 1 & -1 & -1 & -1 \\ 
1 & \omega ^{4} & \omega ^{2} & 1 & \omega ^{4} & \omega ^{2} \\ 
1 & \omega ^{2} & \omega ^{4} & -1 & \omega ^{5} & \omega 
\end{pmatrix},  
\label{E.unitary}
\end{equation}
and $T_{M}({\bf k})$ are the Bloch factors, accounting for the fact
that the 6 sites forming a QMO belong to several different unit
cells. Actual values for these factors depend on the manner in which a
particular band structure code selects the unit cell (see the Appendix
for the WIEN2k settings).

\section{Non-Relativistic electronic structure}
\label{sec:four}

In this Section we analyze and discuss the Ir-Ir $5d$ $t_{2g}$
tight-binding parameters for {\nio} {\it up to second nearest neighbors}.
 As
mentioned in Section~\ref{sec:two}, three structural distortions are
present in {\nio}: orthorhombic distortion, IrO$_6$ octahedra rotation
and trigonal distortion. Besides, the stacking of the honeycomb planes
inherently violates the rhombohedral symmetry even if each plane is
ideal. The formation of QMOs relies on the dominance of
\textit{intra}hexagon hopping~\cite{we}
 and therefore is sensitive to structural
details. Therefore it is important to understand the role of
structural distortions in establishing electron hopping paths. This
motivates us to study electronic properties of a number of
artificially idealized {\nio} unit cells where structural distortions
of different types are systematically eliminated~\cite{Radu_comment}.
Such a procedure has proven very useful~\cite{Martins11} in
understanding the behavior of Sr$_2$IrO$_4$.

We consider four different crystal structures: (i) the experimental
crystal structure~\cite{Radu}, $S_{\rm exp}$, (ii) an artificially idealized {\nio}
unit cell, $S_1$, where the orthorhombic distortion has been removed
from the experimental crystal structure, (iii) an artificially
idealized {\nio} unit cell, $S_2$, where the IrO$_6$ octahedra
rotations have been removed from $S_1$, and (iv) an artificially
idealized {\nio} unit cell, $S_3$, where the trigonal distortion has
been removed from $S_2$.  Table ~\ref{total_energies} 
shows a comparison of total (non-magnetic) DFT energies for the various
structures.
\begin{table}[ptb]
\begin{center}
\begin{tabular}{r|cccc}
Structure &  $S_{\rm exp}$ & $S_1$ & $S_2$ & $S_3$ \\
\hline
$E_{S_i}-E_{S_{\rm exp}}$ (mRyd) & 0 & 0.95 & 78.90 & 180.01
\end{tabular}
\end{center}
\caption{Non-relativistic total energies obtained within DFT for
the experimental, $E_{S_{\rm exp}}$, and the three idealized, $S_i$ ($i=1,2,3$),
{\nio} crystal structures. Energy is given per unit cell containing two formula units.}
\label{total_energies}
\end{table}
As it is to be expected, the experimental structure is the energetically
most stable case.  Tight-binding hopping parameters between Ir $t_{2g}$ orbitals
         up to second nearest neighbors
         calculated for the four structures are given in Table~\ref{T.AOhoppings} and
schematically represented in Fig.~\ref{hoppings}.  We consider the
following rationale for labeling of the hopping parameters
Eq.~\ref{TB_param}.  In the experimental structure of {\nio} there are
two first NN Ir-Ir distances and two second NN Ir-Ir distances due to
the fact that the Ir$_6$ hexagons are not perfect. We denote the
corresponding Ir $t_{2g}$ - Ir $t_{2g}$ hopping parameters as $t_1$
and $t_{\bar{1}}$ for the first NN and, respectively, $t_2$ and
$t_{\bar{2}}$ for the second NN hoppings. Further, we have various
possible hoppings between equal and different $t_{2g}$ orbitals.
Regarding first NN, we denote $t_{1\,{\rm O}}$ and $t_{{\bar{1}}\,{\rm
    O}}$ the hoppings between unlike $t_{2g}$ orbitals via O $p$
states (Fig.~\ref{hoppings}~(a)).  $t_{1\sigma}$ and
$t_{\bar{1}\sigma}$ denote NN direct hoppings of
$\sigma$-type. $t_{1\parallel}$ and $t_{\bar{1}\parallel}$ denote NN
hoppings between like orbitals lying in parallel planes.  In the ideal
structure such hoppings consist of linear combinations with equal
weight of $dd\pi$ and $dd\delta$ bonds. $t_{1\perp}$ and
$t_{\bar{1}\perp}$ denote NN hoppings between unlike orbitals lying in
perpendicular planes (see Figs.~\ref{hoppings}~(b) and (c)).

Regarding the second NN hopping parameters, $t_{2\,{\rm O}}$ and
$t_{\bar{2}\,{\rm O}}$ denote hoppings between unlike orbitals via O
$p$ and Na $s$ states (Fig.~\ref{hoppings}~(e)).  $t_{2a}$ and
$t_{2b}$ ($t_{\bar{2}a}$ and $t_{\bar{2}b}$) denote hoppings between
like orbitals as shown in Fig.~\ref{hoppings}~(d) and $t_{2c}$,
$t_{2d}$ and $t_{2e}$ ($t_{\bar{2}c}$, $t_{\bar{2}d}$ and
$t_{\bar{2}e}$) denote hoppings between unlike orbitals
(Fig.~\ref{hoppings}~(e)).

\subsection{Experimental crystal structure}

Previous electronic structure calculations~\cite{we} have identified
the dominant hopping integrals for {\nio} to be $t_{1\,{\rm O}}$ and
$t_{2\,{\rm O}}$ [as well as $t_{\bar{1}\,{\rm O}}$ and
$t_{\bar{2}\,{\rm O}}$; further on, if not explicitly stated
otherwise, we refer to both equivalent $t_{1}$ ($t_{2}$) and
$t_{\bar{1}}$ ($t_{\bar{2}}$) when writing $t_{1}$ ($t_{2}$)].  In
Table~\ref{T.AOhoppings} column $S_{exp}$ we present the complete list
of hopping parameters up to the second nearest neighbors. A TB model
based  only on these hopping integrals provides
          already a reasonable description of {\nio}
Ir~$t_{2g}$ states near the Fermi level $E_{\text{F}}$
[Fig.~\ref{F.AObands}~(a)].

\begin{figure*}[hbt]
\begin{center}
\includegraphics[width=0.8\textwidth]{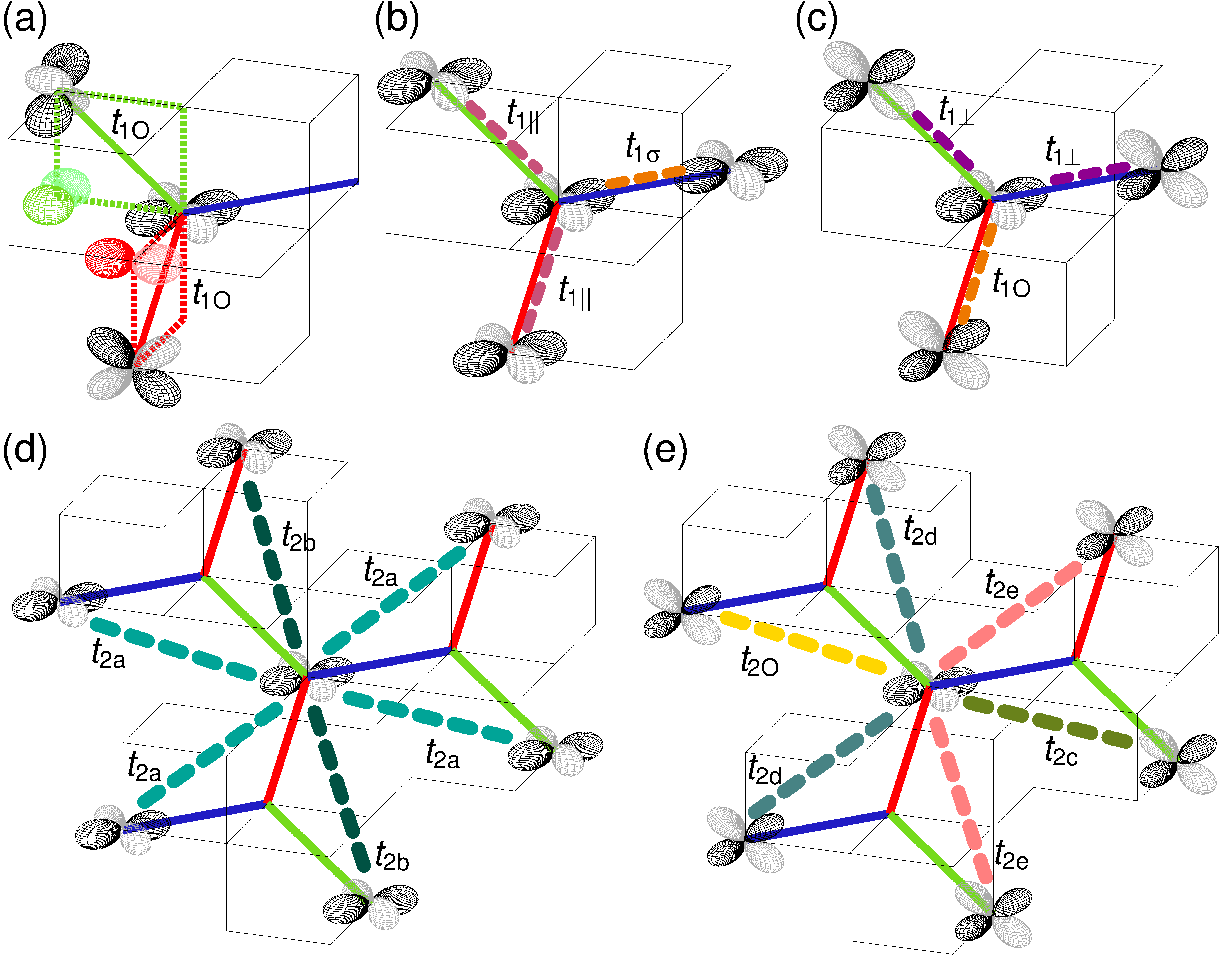}
\end{center}
\caption{Schematic representation of Ir-Ir $t_{2g}$ hopping paths up
  to second nearest neighbor in {\nio}.}
\label{hoppings}
\end{figure*}

We first note a very good agreement between the $t_{1\,{\rm O}}$
($\sim 270$~meV) and $t_{2\,{\rm O}}$ ($\sim -75$~meV) values obtained
with our WIEN2k-based projection method and with the FPLO
code~\cite{FPLO} as was used in Ref.~\onlinecite{we}.  These leading
Ir~$t_{2g}$ hoppings strongly tend to confine the electron's motion to
a single Ir hexagon and, as a result, the electronic structure of
{\nio} near the Fermi level is dominated by the formation of well
separated and relatively weakly dispersive QMOs~\cite{we}. On an Ir
hexagon, as shown above, each Ir atom participates with one of its
$t_{2g}$ orbitals (see Fig. 2 of Ref.~\onlinecite{we}).  These orbitals
combine to form six QMOs according to the unitary transformation
Eq.~(\ref{E.unitary}). In support of this picture,
Fig.~\ref{F.QMOdos}~(a) shows the density of states of {\nio}
projected onto the six QMOs (singlets $A_{1g}$ and $B_{1u}$ and
doublets $E_{2u}$ and $E_{1g}$), where states with certain predominant
QMO character are clearly separated in energy from one another.  The
near-degeneracy of $A_{1g}$ and $E_{2u}$ states around $E_{\text{F}}$
is rather accidental resulting from the $t_{1\,{\rm O}}/t_{2\,{\rm
    O}}\sim-3.6$ ratio  (see Table~\ref{T.AOhoppings} and Ref.~\onlinecite{we}).
 The real-space representations of the QMO
Wannier functions onto which the {\nio} DOS is being projected are
shown in Fig.~\ref{F.Wannier}.  The QMO Wannier functions were
constructed as described in Section~\ref{sec:two} by explicitly
accounting for the location of each Ir~$t_{2g}$ orbital in the
crystal~\cite{note1}.

Other NN and second NN hopping processes involving intraorbital and
interorbital hoppings (see Table~\ref{T.AOhoppings}) allow an electron
to jump from one QMO to another and hence are responsible for the band
dispersion. Many of those hoppings are of the same order of magnitude
(although mostly by at least an order of magnitude smaller) than
$t_{2\,{\rm O}}$, like, for example, $t_{1\parallel}$ and
$t_{\bar{1}\parallel}$.  For the \textquotedblleft
z\textquotedblright\ bond such hoppings will be between $xz$ and $xz$
or $yz$ and $yz$ orbitals (see Fig.~\ref{hoppings} (b)). These
hoppings are equal to 47.7, 30.0, and 33.1~meV, depending on the NN
bond (see Table~\ref{T.AOhoppings}). In fact, such appreciable
variations in magnitude, which violate the $D_{6h}$ symmetry of an
ideal Ir hexagon, are ubiquitous among the hoppings that connect
neighboring QMOs. Some of them even change sign, as, for instance
$t_{1\sigma}$ and $t_{\bar{1}\sigma}$. This feature results from the
orthorhombic stacking, distortions within the Ir$_{2}$Na planes, and
rotations of IrO$_{6}$ octahedra.

\begin{table}[ptb]
  \caption{Nearest neighbor (NN) and second NN
    hopping integrals in meV between Ir~$t_{2g}$ orbitals for the
    experimental structure and three idealized structures 
    $S_1$, $S_2$, $S_3$ of {\nio} (see text and Appendix for a 
    description of the structures and
    parameter labeling). The $\text{NN}=0$ data are Ir~$t_{2g}$ 
    on-site energies and \textit{inter}orbital
    hoppings; the $\text{NN}=1$ and $\text{NN}=\bar{1}$ ($\text{NN}=2$ and 
    $\text{NN}=\bar{2}$) data are hoppings over nonequivalent (due to
    orthorhombic distortion) NN (second NN) Ir bonds.}
\label{T.AOhoppings}
\begin{tabular}{ccrrrr}
\hline\hline
NN &  & $S_{exp}$ & $S_1$ & $S_2$ & $S_3$ \\ \hline
0 & $xy\to xy$ & -448.8 & -422.9 & -422.8 & -601.1 \\ 
& $xz\to xz$ & -421.5 & -421.8 & -421.2 & -601.1 \\ 
& $yz\to yz$ & -421.5 & -421.8 & -421.2 & -601.1 \\ 
& $xy\to xz$, $xy\to yz$ & -27.8 & -26.4 & -21.2 & -13.5 \\ 
& $xz\to yz$ & -23.1 & -25.2 & -18.8 & -14.7 \\ 
&  &  &  &  &  \\ 
1 & $xy\to xy$ ($t_{1\parallel}$) & 47.7 & 34.1 & 27.8 & 120.8 \\ 
& $xy\to xz$, $xy\to yz$ ($t_{1\,{\rm O}}$) & 269.6 & 268.5 & 231.7 & 209.7 \\ 
& $xy\to xz$, $xy\to yz$ ($t_{1\perp}$ ) & -25.6 & -16.6 & 43.7 & -5.3 \\ 
& $xz\to xz$, $yz\to yz$ ($t_{1\parallel}$) & 30.0 & 33.2 & 17.2 & 118.9 \\ 
& $xz\to xz$, $yz\to yz$ ($t_{1\sigma}$) & -20.7 & 3.5 & -66.5 & -381.6 \\ 
& $xz\to yz$ ($t_{1\perp}$) & -21.4 & -16.4 & 41.7 & -4.9 \\ 
&  &  &  &  &  \\ 
$\bar{1}$ & $xy\to xy$ ($t_{\bar{1}\sigma}$) & 25.4 & 0.2 & -65.5 & -382.8 \\ 
& $xy\to xz$, $xy\to yz$ ($t_{\bar{1}\perp}$) & -11.9 & -17.6 & 46.9 & -5.3 \\ 
& $xz\to xz$, $yz\to yz$ ($t_{\bar{1}\parallel}$) & 33.1 & 33.9 & 21.2 & 120.5 \\ 
& $xz\to yz$ ($t_{\bar{1}\,{\rm O}}$) & 264.4 & 264.8 & 228.7 & 211.7 \\ 
&  &  &  &  &  \\ 
2 & $xy\to xy$ ($t_{2a}$) & -3.5 & -2.6 & -18.9 & 2.0 \\ 
& $xy\to xz$, $xy\to yz$ ($t_{2\,{\rm O}}$) & -75.8 & -77.4 & -94.7 & -82.1 \\ 
& $xy\to xz$, $xy\to yz$ ($t_{2c}$) & -36.5 & -35.3 & -52.1 & -38.5 \\ 
& $xy\to xz$, $xy\to yz$ ( $t_{2d}$) & 12.5 & 10.1 & 1.7 & 6.9 \\ 
& $xy\to xz$, $xy\to yz$ ($t_{2e}$) & -21.4 & -19.2 & -7.3 & 1.9 \\ 
& $xz\to xz$, $yz\to yz$ ( $t_{2a}$) & -0.6 & -3.1 & -16.6 & 1.4 \\ 
& $xz\to xz$, $yz\to yz$ ($t_{2b}$) & -1.5 & -1.6 & -1.0 & 5.7 \\ 
& $xz\to yz$ ($t_{2e}$) & -18.6 & -19.0 & -7.1 & 2.4 \\ 
& $xz\to yz$ ($t_{2d}$) & 10.2 & 10.2 & 2.4 & 6.6 \\ 
&  &  &  &  &  \\ 
$\bar{2}$ & $xy\to xy$ ($t_{\bar{2}b}$) & -1.4 & -1.4 & -1.2 & 5.7 \\ 
& $xy\to xz$, $xy\to yz$ ($t_{\bar{2}e}$) & -19.0 & -19.2 & -8.4 & 2.1 \\ 
& $xy\to xz$, $xy\to yz$ ($t_{\bar{2}d}$) & 9.3 & 10.2 & 0.7 & 7.5 \\ 
& $xz\to xz$, $yz\to yz$ ($t_{\bar{2}a}$) & -1.4 & -3.0 & -17.7 & 1.5 \\ 
& $xz\to yz$ ($t_{\bar{2}\,{\rm O}}$) & -77.0 & -78.0 & -95.2 & -81.9 \\ 
& $xz\to yz$ ($t_{\bar{2}c}$) & -30.4 & -35.1 & -51.6 & -38.9 \\ \hline\hline
\end{tabular}
\end{table}

\begin{figure}[ptb]
\begin{center}
\includegraphics[width=0.45\textwidth]{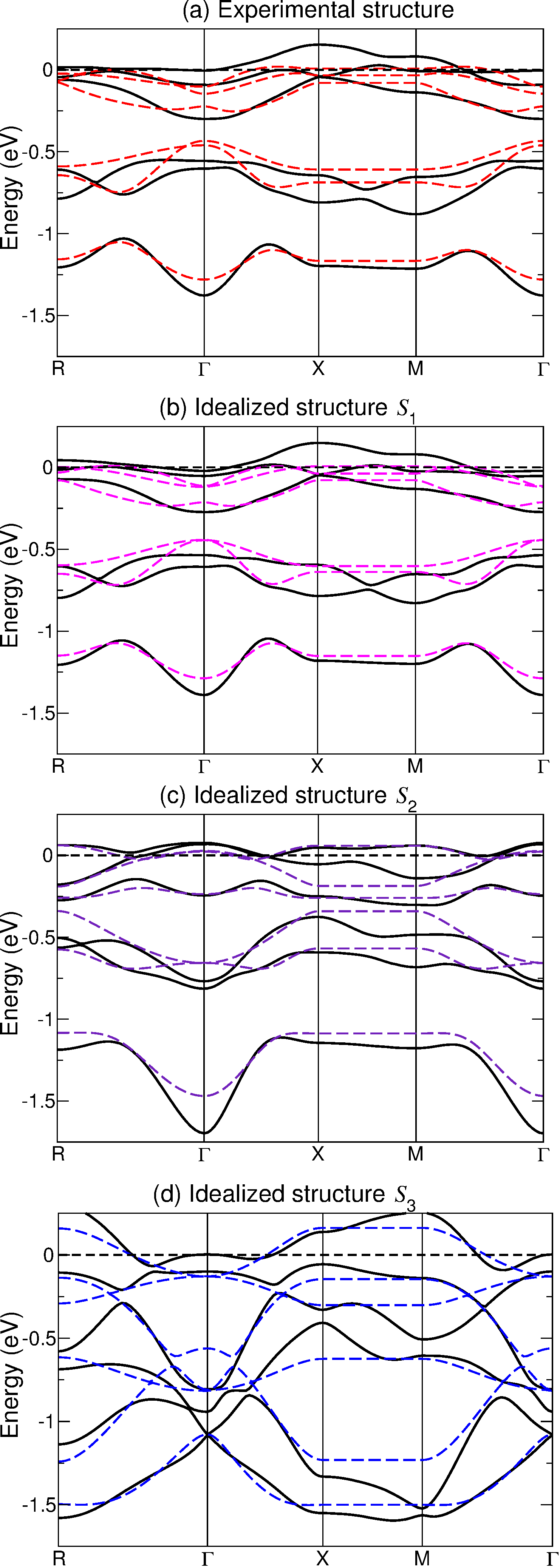}
\end{center}
\caption{{\nio} bandstructures near the Fermi level $E_{\text{F}}=0$
  calculated using DFT (black solid lines) and a TB model that
  considers {\it only} up to NNN hopping processes between Ir~$t_{2g}$ orbitals
  (dashed lines).  The data are obtained with (a) experimental
  crystal structure and idealized structures (b) $S_1$, (c) $S_2$, and
  (d) $S_3$. }
\label{F.AObands}
\end{figure}

\begin{figure}[ptb]
\begin{center}
\includegraphics[width=0.4\textwidth]{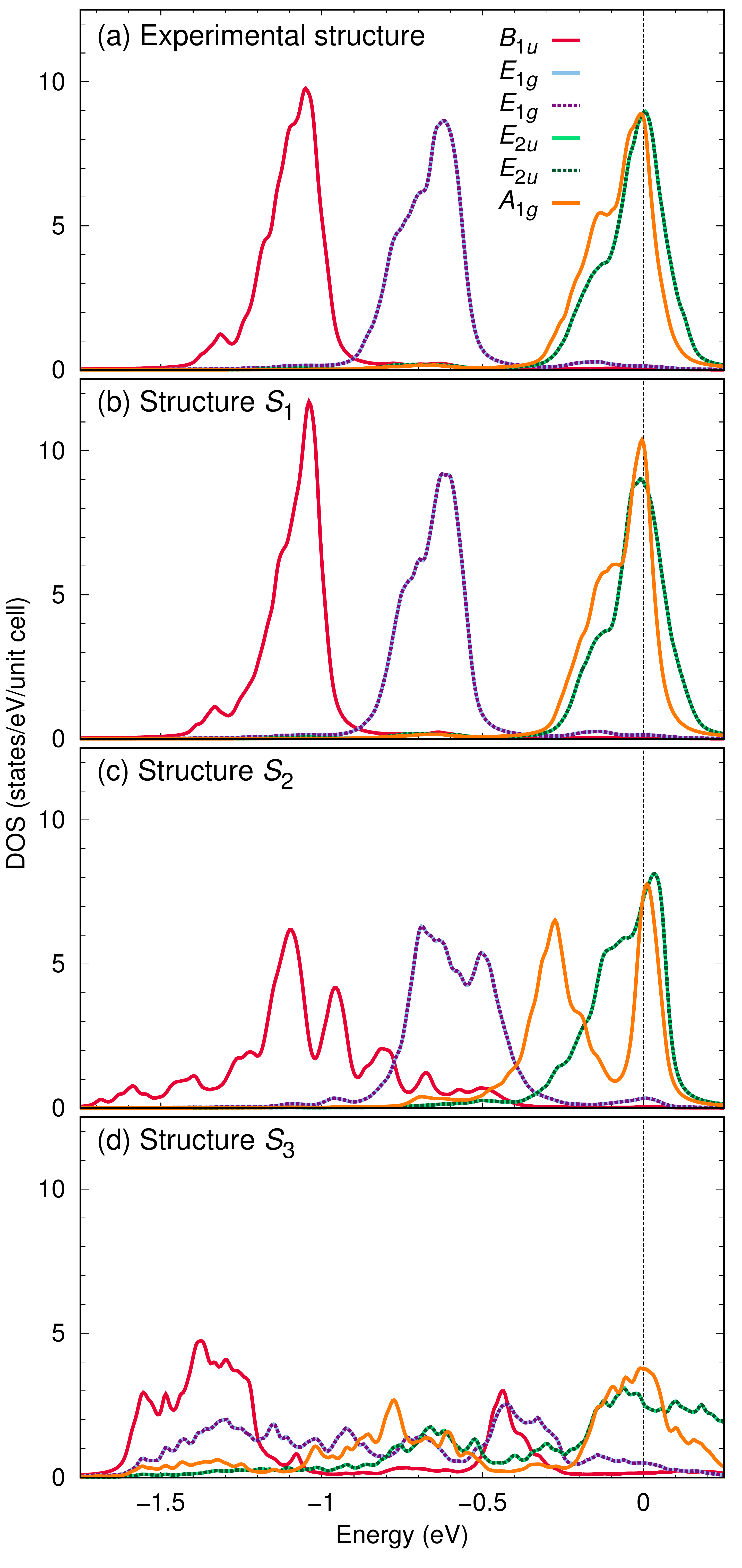}
\end{center}
\caption{{\nio} DOS projected onto QMOs for (a) experimental crystal
  structure and idealized structures (b) $S_1$, (c) $S_2$, and (d)
  $S_3$. The Fermi level is set to zero.}
\label{F.QMOdos}
\end{figure}

\begin{figure}[ptb]
\begin{center}
\includegraphics[width=0.5\textwidth]{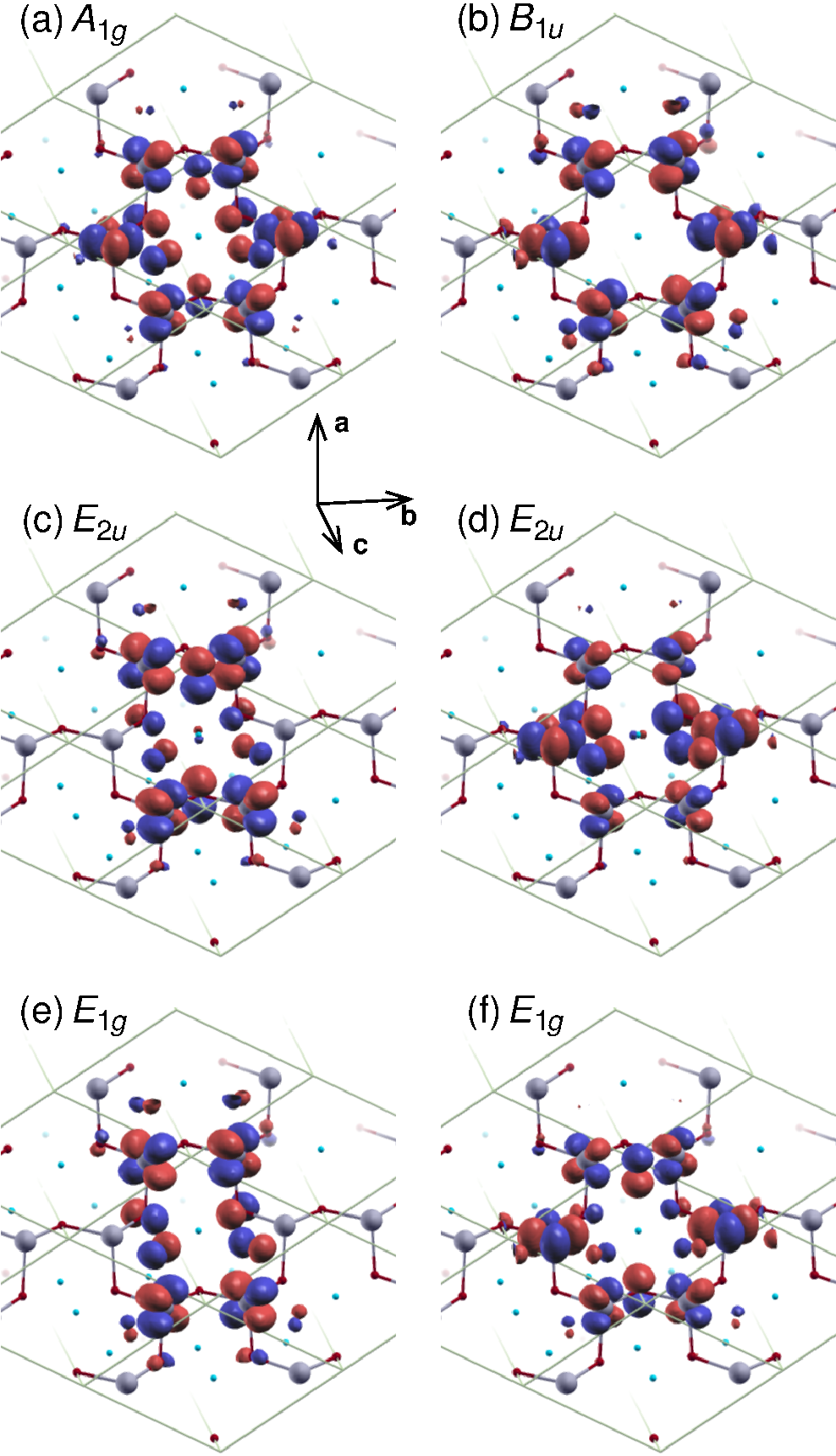}
\end{center}
\caption{Real-space representation of the QMOs in {\nio} obtained by
  the Wannier projector method.}
\label{F.Wannier}
\end{figure}

\subsection{Structure $S_1$ obtained by removing the orthorhombic
  distortion}

We now consider an idealized {\nio} structure without the
orthorhombic distortion of Ir hexagons; this structure, which we call
$S_1$, as well as other structures in this Section, is tabulated in
the Appendix. In the structure $S_1$: (i) all \textit{intra}layer
Ir-Ir bonds are of the same length, {\textit{i.e.}}, the $D_{6h}$
symmetry of an Ir hexagon is restored, (ii) all NN Ir-O bonds are of
the same length, (iii) all Ir-O-Ir bond angles are equal to
98.7$^\circ$, and (iv) the oxygens lie on the faces of a cube drawn
around an Ir hexagon (see Fig.~2 of Ref.~\onlinecite{we}).  The 3D
crystal structure, though, remains orthorhombic in this approximation,
due to the presence of multiple Ir layers. This explains small
residual variations among the nominally equivalent TB model parameters
(Table~\ref{T.AOhoppings}, column $S_1$): {\textit{E.g.}}, comparing
parameters labeled with and without overbar; also, onsite
energies like the $xy$ on-site energy is slightly lower than the
$xz/yz$ on-site energy. However, these variations of $t_{2g}$ orbital
on-site energies, as well as of equivalent hopping integrals, are now
noticeably smaller than in the experimental {\nio} structure.

We conclude that removal of the orthorhombic distortion restores (to a
certain degree) the degeneracy of the Ir~$t_{2g}$ orbitals, but does
not change the hierarchy of hopping integrals. In the structure $S_1$,
the $t_{1\,{\rm O}}$ and $t_{2\,{\rm O}}$ values are close to the
respective values in the experimental {\nio} structure and, as a
consequence, the overall structure of the $t_{2g}$ bands is only
slightly changed [Figs.~\ref{F.AObands}~(b) and \ref{F.QMOdos}~(b)].

\subsection{Structure $S_2$ obtained by removing the
 IrO$_6$ octahedra rotations}

In the structure $S_1$ that we designed in the previous Section, two
types of distortions are still present: (i) trigonal squeezing of
IrO$_6$ octahedra along the (111) direction perpendicular to Ir
hexagon planes and (ii) IrO$_{6}$ octahedra rotations that place O
atoms on the cube's faces. We now consider structure $S_2$, where the
IrO$_{6}$ octahedra rotations are removed from $S_1$.  In this
structure, the Na-O and Ir-O bond lengths are the same (in the
experimental structure, the former is considerably longer). This
feature enhances the second NN hopping processes through Na~$s$
states, such as $t_{2\,{\rm O}}$, $t_{2a}$, $t_{2c}$ (and the
equivalent overbar hoppings) as shown in Table~\ref{T.AOhoppings},
column $S_2$.  At the same time, the NN O-assisted hopping $t_{1\,{\rm
    O}}$ gets reduced and the $t_{1\,{\rm O}}/t_{2\,{\rm O}}$ ratio
decreases to $\sim -2.4$, resulting in a
larger separation of the lowest ($B_{1u}$) band from the rest of $t_{2g}$
bands [Fig.~\ref{F.AObands}~(c)].
Formation of QMOs still takes place in
structure $S_2$ [Fig.~\ref{F.QMOdos}~(c)], but the QMO bands are more
dispersive compared to the experimental or $S_1$ structures, due to
increased \textit{inter}hexagon NN hopping integrals $t_{1\sigma}$,
$t_{1\perp}$ (and equivalent $t_{\bar{1}\sigma}$, $t_{\bar{1}\perp}$): thus,
one observes broadening of the $A_{1g}$ band and redistribution of weight away from the
$E_{2u}$ doublet.

\subsection{Structure $S_3$ obtained by removing
the  trigonal distortion}

We finally consider a most idealized {\nio} structure $S_3$ without
the trigonal distortion, {\textit{i.e.}}, with 90$^\circ$~Ir-O-Ir
bond angles.  Importantly, one can only remove this distortion, while
keeping the Ir-O bond length the same, if the Ir-Ir bonds are
shortened. Because of that, the hierarchy of hopping integrals changes
drastically (Table~\ref{T.AOhoppings}, column $S_3$).  The dominant
hopping is now the direct NN hopping between like orbitals
$t_{1\sigma}$ (and the equivalent $t_{\bar{1}\sigma}$) reaching $\sim
-380$~meV, while the O-assisted hopping $t_{1\,{\rm O}}$
($t_{\bar{1}\,{\rm O}}$) has been reduced to $\sim 210$~meV.
Accordingly, the large \textit{inter}hexagon interaction destroys the
QMO picture, as illustrated by the strongly dispersive $t_{2g}$
manifold in Fig.~\ref{F.AObands}~(d) and the delocalization of
individual QMO characters over the whole DOS range in
Fig.~\ref{F.QMOdos}~(d).  We also observe that the main reason for the
trigonal squeeze is the geometrical effect of optimizing
simultaneously the Ir-Ir and Ir-O bonds. As a result, even though the
on-site $t_{2g}$ orbitals split into an $a_{1g}$ singlet and an
$e_{g}$ doublet, this is not a strong effect and not the driving force
for the squeeze, as it is often assumed in the spirit of localized
limit and the Jahn-Teller effect.

Summarizing these results, in the $S_3$ structure, the NN direct
hopping increases by an order of magnitude compared to the
experimental {\nio} structure and the NN O-assisted hoppings get
suppressed. Therefore we conclude that structural distortions of all
types in {\nio} act constructively to enhance the
\textit{intra}hexagon effective hopping parameters (such as
$t_{1\,{\rm O}}$ and $t_{2\,{\rm O}}$ ) and suppress the
\textit{inter}hexagon ones (such as NN direct hopping) favoring the
formation of QMOs.

\section{Spin-orbit coupling}
\label{sec:five}

We proceed now with the analysis of the electronic structure of {\nio}
in the presence of spin-orbit (SO) coupling. Previous relativistic DFT
calculations~\cite{Shitade2009} showed that {\nio} states near the
Fermi level experience strong relativistic splitting with pronounced
concentration of $j_{\text{eff}}=\frac{1}{2}$ character in the upper
two bands. However, the {\nio} relativistic states seem to preserve
their QMO identity as well [see Fig.~S6~(b) of
Ref.~\onlinecite{we}]. In order to understand such duality, we set up
a TB model for the Ir~$t_{2g}$ orbitals that includes also local SO
interaction terms.  With this TB+SO model, we are able not only to
confirm the relativistic DFT results by calculating DOS but also to access the
composition of individual states and trace their evolution as a function
of the spin-orbit coupling $\lambda$.

\subsection{TB+SO model}
\label{sec:fiveA}

We start with a TB model that perfectly describes the non-relativistic
DFT Ir~$t_{2g}$ bands of {\nio}. It includes three hundred and twenty
one hopping integrals between up to 50 nearest neighbors. We then
double the dimension of the TB Hamiltonian matrix to introduce spin
dependence and add local SO coupling terms $\langle \lambda
{\bf L}\cdot {\bf S}\rangle $ that mix spin-$\uparrow $ and
spin-$\downarrow $ subspaces:
\begin{equation}
\begin{array}{c|cccccc}
& {xy}\uparrow  & {xz}\uparrow  & {yz}\uparrow  & {xy}\downarrow  & 
{xz}\downarrow  & {yz}\downarrow  \\ \hline
{xy}\uparrow  & 0 & 0 & 0 & 0 & \frac{\lambda }{2} & -\frac{i\lambda }{2}
\\ 
{xz}\uparrow  & 0 & 0 & \frac{i\lambda }{2} & -\frac{\lambda }{2} & 0 & 0
\\ 
{yz}\uparrow  & 0 & -\frac{i\lambda }{2} & 0 & \frac{i\lambda }{2} & 0 & 0
\\ 
{xy}\downarrow  & 0 & -\frac{\lambda }{2} & -\frac{i\lambda }{2} & 0 & 0 & 
0 \\ 
{xz}\downarrow  & \frac{\lambda }{2} & 0 & 0 & 0 & 0 & -\frac{i\lambda }{2}
\\ 
{yz}\downarrow  & \frac{i\lambda }{2} & 0 & 0 & 0 & \frac{i\lambda }{2} & 0
\end{array}\label{matrix_SO}
\end{equation}
Importantly, even though SO coupling is a local on-site interaction,
it couples neighboring quasi-molecular orbitals and therefore
is {${\bf k}$-vector} dependent in the QMO basis.

\begin{figure}[tb]
\includegraphics[width=0.45\textwidth]{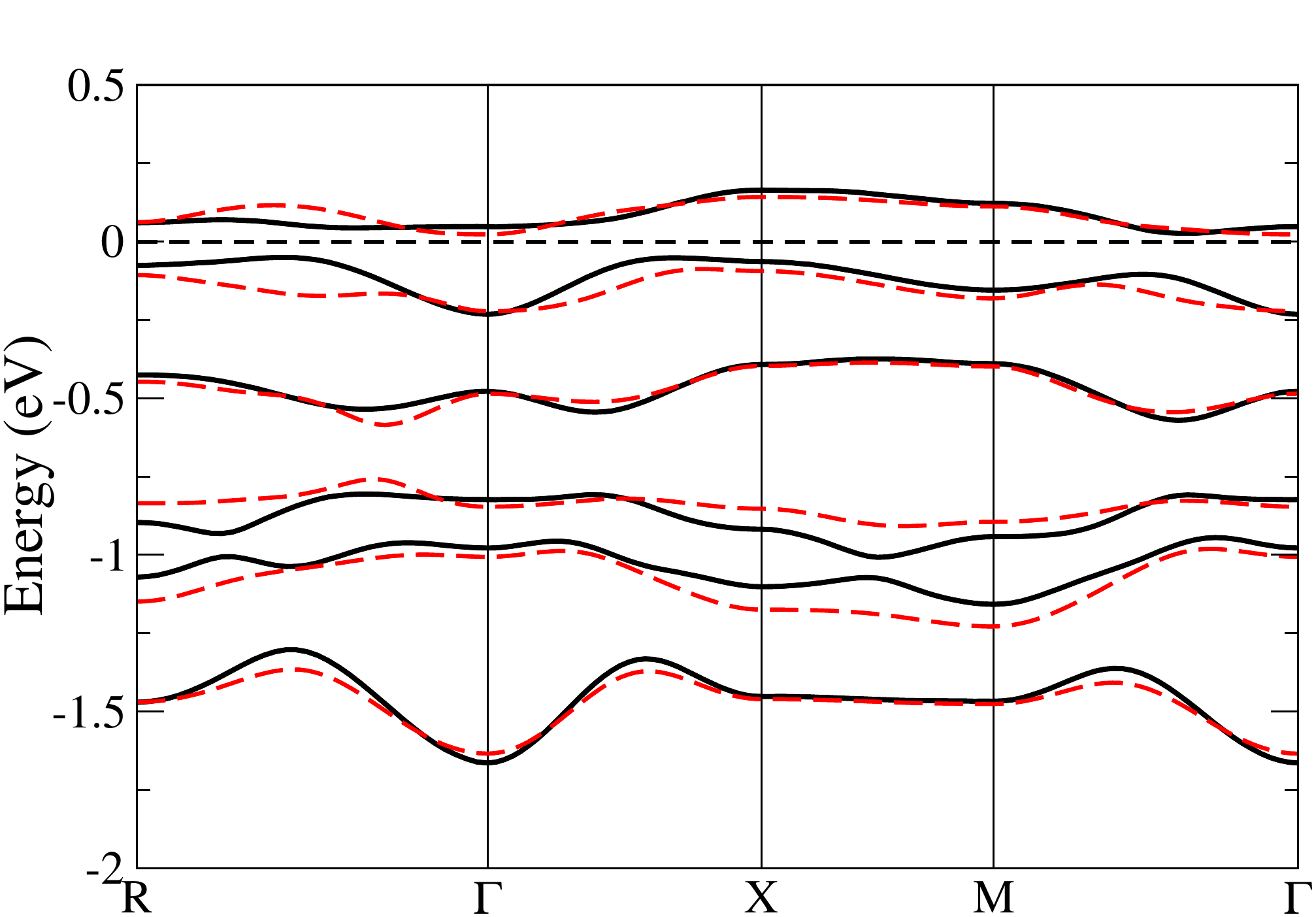}
\caption{WIEN2k relativistic bandstructure (black solid lines) versus
  TB+SO model relativistic bandstructure (red dashed lines) of {\nio}
  as described in the text. In model calculations,
  $\protect\lambda=0.44$~eV was used.}
\label{F.RELbands}
\end{figure}

Having thus set up the TB model, we vary the SO coupling strength
$\lambda$ until the best matching with the DFT relativistic bands is
achieved, which is found to correspond to $\lambda=0.44$~eV
(Fig.~\ref{F.RELbands}).

Since our purpose is to reconcile the QMO and relativistic orbital
(RO) pictures, we analyze the $\lambda{\bf L}\cdot{\bf S}$ matrix
elements between spin-$\uparrow$ and spin-$\downarrow$ QMOs to see how
SO coupling mixes QMO characters. They can be easily obtained by
applying the unitary transformation $UT({\bf k})$
[Eq. (\ref{E.unitary}) ] to the $\lambda{\bf L}\cdot{\bf S}$ matrix
in the $t_{2g}$ basis:
\begin{equation}
H^{\text{SO}}_{\text{QMO}}({\bf k}) = U T({\bf k}) H^{\text{SO}}_{t_{2g}}
T^{H}({\bf k}) U^{H}.   \label{E.SOtrans}
\end{equation}
This equation explicitly illustrates how {${\bf k}$-vector} dependence
enters the SO matrix elements in the QMO basis. Concise expressions
can be derived if one notes that QMOs can be represented by their
``winding number'' $n$ which defines a phase change
$\Delta\phi=\frac{n\pi}{3}$ of $t_{2g}$ orbitals around a hexagon. In
this notation, QMOs $A_{1g}, E_{2u}, E_{1g}, B_{1u}, E_{1g}, E_{2u}$
correspond to, respectively, $n=0,1,2,3,4,5$ winding numbers. The
$\lambda{\bf L}\cdot{\bf S}$ matrix elements in the QMO basis are
given then by
\begin{align}
H^{\text{SO}}_{n\uparrow n^{\prime}\uparrow} & = \frac{\lambda}{2} ie^{\frac{(n^{\prime}-n)\pi i}{2}} \cos\frac{(n^{\prime}-n)\pi}{2}\cos (k_{x}+k_{y}) 
\notag \\
& \times\left( e^{\frac{2(2n^{\prime}-n)\pi i}{3}}-e^{-\frac{2(2n-n^{\prime
})\pi i}{3}}\right)  \notag \\
& + \frac{\lambda}{2} ie^{\frac{(n^{\prime}-n)\pi i}{2}} \sin\frac{(n^{\prime }-n)\pi}{2}\sin(k_{x}+k_{y})  \notag \\
& \times\left( e^{\frac{2(2n^{\prime}-n)\pi i}{3}}+e^{-\frac{2(2n-n^{\prime
})\pi i}{3}}\right)
\end{align}
and 
\begin{align}
H^{\text{SO}}_{n\uparrow n^{\prime}\downarrow} & = 2\,e^{\frac{(n^{\prime
}-n)\pi i}{2}} \left( e^{\frac{4n^{\prime}\pi i}{3}}\cos(-\frac{(n^{\prime
}-n)\pi}{2} + k_{y})\right.  \notag \\
& -e^{-\frac{4n\pi i}{3}}\cos(-\frac{(n^{\prime}-n)\pi}{2} - k_{y})  \notag
\\
& + ie^{\frac{2n^{\prime}\pi i}{3}}\cos(-\frac{(n^{\prime}-n)\pi}{2} - k_{x})
\notag \\
& \left. - ie^{-\frac{2n\pi i}{3}}\cos(-\frac{(n^{\prime}-n)\pi}{2} + k_{x})
\right) .
\end{align}
We list numerical values of the matrix elements for two representative
{${\bf k}$-vector}s: ${\bf k}=(0,0,0)$ (point $\Gamma$) (first two
tables) and ${\bf k}=(\frac{\pi}{2},0,0)$ (last two tables).

\begin{equation}  \label{E.Guu}
\begin{array}{c|ccc|ccc}
& n=5 & n=0 & n=1 & n=2 & n=3 & n=4 \\ 
& E_{2u}\uparrow & A_{1g}\uparrow & E_{2u}\uparrow & E_{1g}\uparrow & 
B_{1u}\uparrow & E_{1g}\uparrow \\ \hline
E_{2u}\uparrow & C_{1} & 0 & 0 & 0 & -C_{1} & 0 \\ 
A_{1g}\uparrow & 0 & 0 & 0 & -C_{1} & 0 & C_{1} \\ 
E_{2u}\uparrow & 0 & 0 & -C_{1} & 0 & C_{1} & 0 \\ \hline
E_{1g}\uparrow & 0 & -C_{1} & 0 & C_{1} & 0 & 0 \\ 
B_{1u}\uparrow & -C_{1} & 0 & C_{1} & 0 & 0 & 0 \\ 
E_{1g}\uparrow & 0 & C_{1} & 0 & 0 & 0 & -C_{1}
\end{array}
\end{equation}
\begin{equation}  \label{E.Gud}
\begin{array}{c|ccc|ccc}
& n=5 & n=0 & n=1 & n=2 & n=3 & n=4 \\ 
& E_{2u}\downarrow & A_{1g}\downarrow & E_{2u}\downarrow & E_{1g}\downarrow
& B_{1u}\downarrow & E_{1g}\downarrow \\ \hline
E_{2u}\uparrow & C_{2} & 0 & 0 & 0 & C_{4} & 0 \\ 
A_{1g}\uparrow & 0 & 0 & 0 & -C_{3} & 0 & -C_{4} \\ 
E_{2u}\uparrow & 0 & 0 & -C_{2} & 0 & C_{3} & 0 \\ \hline
E_{1g}\uparrow & 0 & C_{4} & 0 & C_{2} & 0 & 0 \\ 
B_{1u}\uparrow & -C_{3} & 0 & -C_{4} & 0 & 0 & 0 \\ 
E_{1g}\uparrow & 0 & C_{3} & 0 & 0 & 0 & -C_{2}
\end{array}
\end{equation}
\begin{equation}
\begin{array}{c|ccc|ccc}
& n=5 & n=0 & n=1 & n=2 & n=3 & n=4 \\ 
& E_{2u}\uparrow & A_{1g}\uparrow & E_{2u}\uparrow & E_{1g}\uparrow & 
B_{1u}\uparrow & E_{1g}\uparrow \\ \hline
E_{2u}\uparrow & \frac{\lambda}{6} & 0 & \frac{\lambda}{6} & 0 & -\frac{\lambda}{3} & 0 \\ 
A_{1g}\uparrow & 0 & \frac{\lambda}{6} & 0 & -\frac{\lambda}{3} & 0 & \frac{\lambda}{6} \\ 
E_{2u}\uparrow & \frac{\lambda}{6} & 0 & -\frac{\lambda}{3} & 0 & \frac{\lambda}{6} & 0 \\ \hline
E_{1g}\uparrow & 0 & -\frac{\lambda}{3} & 0 & \frac{\lambda}{6} & 0 & \frac{\lambda}{6} \\ 
B_{1u}\uparrow & -\frac{\lambda}{3} & 0 & \frac{\lambda}{6} & 0 & \frac{\lambda}{6} & 0 \\ 
E_{1g}\uparrow & 0 & \frac{\lambda}{6} & 0 & \frac{\lambda}{6} & 0 & -\frac{\lambda}{3}
\end{array}
\end{equation}
\begin{equation}
\begin{array}{c|ccc|ccc}
& n=5 & n=0 & n=1 & n=2 & n=3 & n=4 \\ 
& E_{2u}\downarrow & A_{1g}\downarrow & E_{2u}\downarrow & E_{1g}\downarrow
& B_{1u}\downarrow & E_{1g}\downarrow \\ \hline
E_{2u}\uparrow & C_{1} i & C_{5} & 0 & -\frac{\lambda}{6} & C_{6} & C_{7} \\ 
A_{1g}\uparrow & C^{*}_{5} & 0 & C_{5} & -C^{*}_{6} & \frac{\lambda}{3} & 
-C_{6} \\ 
E_{2u}\uparrow & 0 & C^{*}_{5} & -C_{1} i & C^{*}_{7} & C^{*}_{6} & -\frac{\lambda}{6} \\ \hline
E_{1g}\uparrow & -\frac{\lambda}{6} & C_{6} & C_{7} & C_{1} i & C_{5} & 0 \\ 
B_{1u}\uparrow & -C^{*}_{6} & \frac{\lambda}{3} & -C_{6} & C^{*}_{5} & 0 & 
C_{5} \\ 
E_{1g}\uparrow & C^{*}_{7} & C^{*}_{6} & -\frac{\lambda}{6} & 0 & C^{*}_{5}
& -C_{1} i\end{array}
\end{equation}
with $C_{1} = \frac{\lambda}{\sqrt{12}}$, $C_{2} = \frac{\lambda}{\sqrt{12}}(1+i)$, $C_{3} = 0.105663\lambda(1+i)$, $C_{4} = 0.394337\lambda(1+i)$, $C_{5} = \frac{\lambda}{12} + \frac{\lambda}{2\sqrt{12}}i$, $C_{6} = \frac{\lambda}{4} + \frac{\lambda}{2\sqrt{12}}i$, $C_{7} = -\frac{\lambda}{6} + 
\frac{\lambda}{\sqrt{12}}i$.

Several comments are in place here. First, spin-orbit coupling mixes QMOs at
all {${\bf k}$-vector}s. Even at the $\Gamma $ point,
{\textit{i.~e.}}, on the same hexagon, the three upper QMOs ($A_{1g}$
and two $E_{2u}$) are  SO coupled to the three lower QMOs ($B_{1u}$ and
two $E_{1g}$), which explains sizable shifts of the relativistic bands
compared to the non-relativistic ones at this
{${\bf k}$-vector}. Additionally, SO coupling induces splitting of the
degenerate $E_{2u}$ and $E_{1g}$ states at all {${\bf k}$-vector}s.
Another striking feature of the calculated $\lambda {\bf L}\cdot
{\bf S}$ matrix is that its $A_{1g}$, $E_{2u}$ (upper triplet) and
$B_{1u}$, $E_{1g}$ (lower triplet) blocks are identical. This means
that if not for the accidental near-degeneracy of the $A_{1g}$ and
$E_{2u}$ states (which magnifies the SO induced energy shifts) the
upper and the lower triplets would have been equally affected by the
SO coupling.

\subsection{Quasimolecular orbital basis versus relativistic basis}

The main difficulty in describing the {\nio} bandstructure is that it
interpolates between eigenstates of two Hamiltonians: the
\textit{itinerant} TB Hamiltonian of (primarily) \textit{intra}hexagon
electron hopping that preserves the $s_z$ spin subspace and the
\textit{local} spin-orbit (SO) interaction
$\lambda{\bf L}\cdot{\bf S}$ Hamiltonian that couples
\textit{different} spin subspaces. The eigenstates of the TB
Hamiltonian are quasi-molecular orbitals (QMOs), while the
eigenstates of the SO interaction (in the $t_{2g}$ subspace) are
\textit{relativistic orbitals} (ROs)
$|j_{\text{eff}},j^{z}_{\text{eff}}\rangle$ characterized by an
effective total angular momentum $j_{\text{eff}}$ and its
$z$-projection $j^{z}_{\text{eff}}$:
\begin{align}
|\tfrac{1}{2},\tfrac{1}{2}\rangle & = \frac{1}{\sqrt{3}}|{xy}\uparrow
\rangle+\frac{i}{\sqrt{3}}|{xz}\downarrow\rangle+ \frac{1}{\sqrt{3}}|{yz}\downarrow\rangle,  \notag \\
|\tfrac{1}{2},-\tfrac{1}{2}\rangle & = \frac{i}{\sqrt{3}}|{xz}\uparrow\rangle-\frac{1}{\sqrt{3}}|{yz}\uparrow\rangle+ \frac{1}{\sqrt{3}}|{xy}\downarrow\rangle,  \notag \\
|\tfrac{3}{2},\tfrac{3}{2}\rangle & = \frac{i}{\sqrt{2}}|{xz}\uparrow
\rangle+\frac{1}{\sqrt{2}}|{yz}\uparrow\rangle,  \notag \\
|\tfrac{3}{2},\tfrac{1}{2}\rangle & = -\sqrt{\frac{2}{3}}|{xy}\uparrow\rangle+\frac{i}{\sqrt{6}}|{xz}\downarrow\rangle+ \frac{1}{\sqrt{6}}|{yz}\downarrow\rangle,  \notag \\
|\tfrac{3}{2},-\tfrac{1}{2}\rangle & = \frac{i}{\sqrt{6}}|{xz}\uparrow\rangle-\frac{1}{\sqrt{6}}|{yz}\uparrow\rangle-\sqrt{\frac{2}{3}}|{xy}\downarrow\rangle,  \notag \\
|\tfrac{3}{2},-\tfrac{3}{2}\rangle & = -\frac{i}{\sqrt{2}}|{xz}\downarrow\rangle+\frac{1}{\sqrt{2}}|{yz}\downarrow\rangle.
\end{align}

This basis~\cite{comment_basis} can be explained as follows; three
$t_{2g}$ orbitals (total degeneracy, including spins, is 6) are split
into a lower-lying quartet $j_{\text{eff}} =3/2$ and an upper lying
$j_{\text{eff}}=1/2$ doublet, and the $5d$-electrons of Ir$^{4+}$
fully occupy the lower quartet leaving the upper $j_{\text{eff}} =1/2$
doublet half-filled. This makes this situation similar to a
non-degenerate Hubbard model (S=1/2 doublet on a site), with the
important difference that in the Hubbard model the hopping matrix
elements preserve the $s_z$ spin subspace, while here the states of
the $j_{\text{eff}}=1/2$ doublet are spin-orbit mixed states, leading
to a strong anisotropy of hoppings and their dependence on spin (or
rather total moment) direction. This may bring about anisotropic
exchange, {\textit{e.~g.}}, the Kitaev exchange on a honeycomb
lattice~\cite{Chaloupka2010}.

\begin{figure}[ptb]
\includegraphics[width=0.45\textwidth]{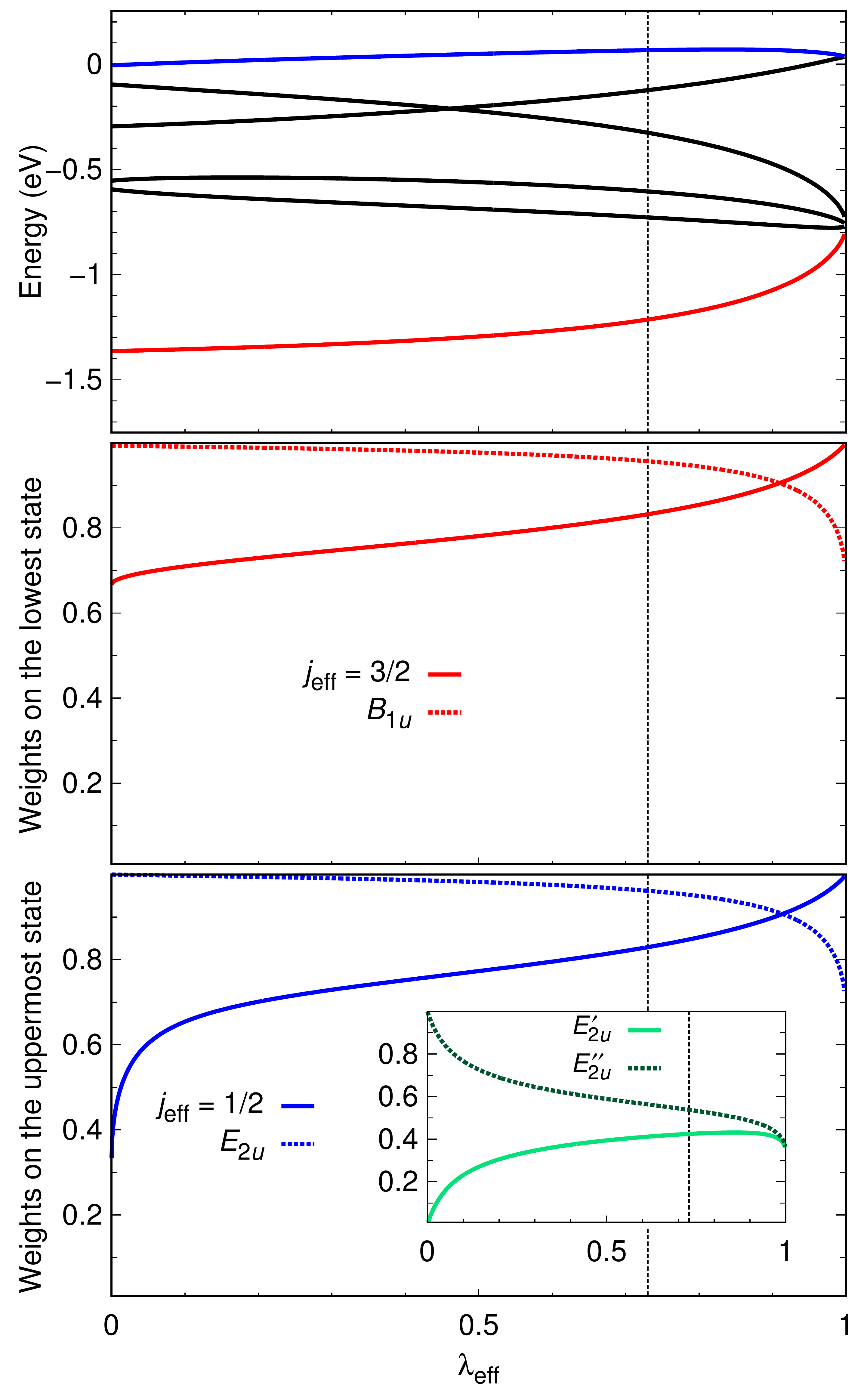}
\caption{Properties of the TB+SO model of {\nio} at the $\Gamma$ point
  as a function of effective SO coupling
  $\protect\lambda_{\text{eff}}$ defined in
  Eq.~(\protect\ref{leff}). The vertical dotted line marks the
  realistic $\protect\lambda_{\text{eff}}=0.73$ value for {\nio}.  (a)
  The eigenvalues of the TB+SO model at $\Gamma$. Eigenenergies have been scaled by
$\sqrt{1-\lambda_{\rm eff}}$ to keep them
  within the $[-1.5,0.2]$~eV range. (b) The
  $j_{\text{eff}}=\frac{3}{2}$ (solid line) and $B_{1u}$ (dashed line)
  weights on the lowest state. (c) The $j_{\text{eff}}=\frac{1}{2}$
  (solid line) and total $E_{2u}$ (dashed line) weights on the
  uppermost state. Inset shows individual contributions from the two
  $E_{2u}$ QMOs. }
\label{F.evolution}
\end{figure}

By gradually increasing an effective spin-orbit coupling strength
$\lambda_{\text{eff}}$,
\begin{equation}
\lambda_{\text{eff}}=\frac{\lambda^{2}}{(
t_{1\,{\rm O}})^{2}+\lambda^{2}},\quad  t_{1\,{\rm O}} =0.270~\mbox{eV},   \label{leff}
\end{equation}
from 0 to 1, one can trace a smooth evolution of the TB+SO model
eigenvalues from, respectively, the non-relativistic (QMO) limit to
the fully relativistic (RO) limit (see Fig.~\ref{F.evolution}~(a) for
the data at the $\Gamma$ point). An SO coupling parameter of
$\lambda=0.44$~eV for {\nio} corresponds to
$\lambda_{\text{eff}}=0.73$, which is marked by a vertical dotted line
in Fig.~\ref{F.evolution}.

The RO basis is an attractive starting point to describe the
low-energy physics of {\nio} as it allows to truncate the Hamiltonian
to only $j_{\text{eff}}=\frac{1}{2}$ states that dominate near the
Fermi energy and map {\nio} onto the Kitaev-Heisenberg model. Although
this approach might seem reasonable given the noticeable separation of
the $j_{\text{eff}}=\frac{1}{2}$ and $j_{\text{eff}}=\frac{3}{2}$
characters in the DOS of {\nio} [{\textit{cf.}} Fig.~2~(b) of
Ref.~\onlinecite{Shitade2009}], we argue that the itinerant terms are
too strong to be neglected (which should not be surprising since
$\lambda =0.44~\mbox{eV}<W\approx 4t_{1\,{\rm O}} =1~\mbox{eV}$) and
that, consequently, the QMO basis is as well (or as poorly) justified
to work with as the RO basis.

To support this statement, let us concentrate on the TB+SO model
states at the $\Gamma$ point.  Fig.~\ref{F.evolution}~(a) shows the
evolution of the model eigenvalues as a function of
$\lambda_{\text{eff}}$ (Eq.~\ref{leff}).  In the non-relativistic limit
($\lambda_{\text{eff}}=0$), the states are almost purely (with slight
deviation due to orthorhombic distortion) QMOs, ordered as
$B_{1u},E_{1g},A_{1g},E_{2u}$ with increasing energy~\cite{note2}.  At
the same time, at each state the $j_{\text{eff}}=\frac{1}{2}$
contribution is $1/3$ and the $j_{\text{eff}}=\frac{3}{2}$
contribution is, correspondingly, $2/3$ (for one of the two Ir
atoms). Note that, since the model distinguishes spin-$\uparrow$ and
spin-$\downarrow$ states each level is doubly degenerate.

With the QMO splitting obviously prevailing for zero SO coupling, we
now want to quantify the QMO character rectification upon increasing
$\lambda_{\text{eff}}$ by calculating the QMO and RO weights on two
selected states: the lowest ($B_{1u}$) and the uppermost
($E_{2u}$). The $B_{1u}$ state [Fig.~\ref{F.evolution}~(b)] is a
simpler case as it is non-degenerate (apart from spin) and quite well
separated from the rest of the QMOs so that the SO effects here should be less
important. Changing $\lambda_{\text{eff}}$ from 0 to 0.73 ({\nio}
value), the $j_{\text{eff}}=\frac{3}{2}$ weight on this state
increases from 0.6667 to 0.8320, whereas the $B_{1u}$ weight is only
slightly reduced from 0.9932 to 0.9567. This indicates that the lowest
relativistic state at the $\Gamma$ point in {\nio} is better described
by a QMO $B_{1u}$ than by one of the $j_{\text{eff}}=\frac{3}{2}$
ROs. In fact, this turns out to hold for  the whole lowest relativistic
band [{\textit{cf.}} the $j_{\text{eff}}$- and QMO-projected {\nio}
DOS in, respectively, Fig.~2~(b) of Ref.~\onlinecite{Shitade2009} and
Fig.~S6~(b) of Ref.~\onlinecite{we}].

The uppermost state is one of the $E_{2u}$ doublet states. It is
near-degenerate with $A_{1g}$ and the other $E_{2u}$ and, therefore,
the SO effects are here particularly strong. At the $\Gamma$ point,
though, it can only couple to itself or to the other $E_{2u}$ [see
Eqs.~(\ref{E.Guu}) and (\ref{E.Gud})], depending on which linear
combination of these degenerate states is considered. Upon switching
$\lambda_{\text{eff}}$ on, the $j_{\text{eff}}=\frac{1}{2}$ weight on
this upper states rapidly grows from 0.3333 to $\sim0.6$ in the range
$0<\lambda_{\text{eff}}<0.05$, and then gradually increases to 0.8295
at $\lambda_{\text{eff}}=0.73$ [Fig.~\ref{F.evolution}~(c)]. At the
same time, the weight of one of the $E_{2u}$ states (we may call it
$E_{2u}^{\prime}$) is reduced from 1.0 to 0.53730 [see inset of
Fig.~\ref{F.evolution}~(c)]. However, the total weight of two $E_{2u}$
states is barely changed: at $\lambda_{\text{eff}}=0.73$ it equals
0.9617. This means that the uppermost relativistic state at the
$\Gamma$ point in {\nio} is very well described by a linear
combination of two $E_{2u}$ states (which is also a QMO) with, in
general, $\lambda_{\text{eff}}$-dependent individual contributions.

The $B_{1u}$ and $E_{2u}$ states (at $\lambda =0$) 
seem to simultaneously bear both RO
and QMO features up to very strong SO coupling, with the QMO character 
dominating for $\lambda_{\text{eff}}<0.9$. This can also be illustrated
by inspecting the composition of, {\textit{e.~g.}}, the lowest
energy band state as shown in Table~\ref{T.states}.
 At zero SO coupling, the doubly
degenerate lowest state corresponds to (almost) pure $B_{1u}\uparrow$
and $B_{1u}\downarrow$ QMOs~\cite{note3}.  At
$\lambda_{\text{eff}}=0.73$, the structure of this state is strikingly
similar to the $B_{1u}$ states,
 with only slight admixtures of the ${xz}$ and ${yz}$ orbitals
of opposite spin. Even at some very high $\lambda _{\text{eff}}$, when
the RO $j_{\text{eff}}=\frac{3}{2}$ weight is close to 1, the states
retain the $B_{1u}\uparrow$ and $B_{1u}\downarrow$ QMO features.
\begin{table*}[tb]
  \caption{Expansion coefficients of  the lowest doubly degenerate
    energy states of the TB+SO model in the $t_{2g}$ basis
(The upper index of the $t_{2g}$
    orbitals labels Ir atoms in the unit cell).
The  coefficients are given for three  $\lambda$ ($\lambda_{\rm eff}$)
values.  The $B_{1u}$ and 
 $j_{\text{eff}}=\frac{3}{2}$ weights of the various states are given
at the bottom of the table.}
\label{T.states}
\begin{center}
\begin{tabular}{ccrrcrrcrr}
\hline\hline
& \hspace{0.5cm}\, & \multicolumn{2}{c}{$\lambda=0$ ($\lambda_{\text{eff}}=0$}) & \hspace{0.5cm}\, &
\multicolumn{2}{c}{$\lambda=0.44$~eV ($\lambda_{\rm eff}=0.73$)} & \hspace{0.5cm}\, &
\multicolumn{2}{c}{$\lambda =2.66$~eV ($\lambda_{\text{eff}}=0.99$)} \\ \hline
${xy}^{1} \uparrow$ &  & -0.454 & {0.0\hspace{0.27cm}\,} &  & -0.453\hspace{1.33cm} & {0.0\hspace{0.27cm}\,}\hspace{1.33cm} &  & -0.444\hspace{1.33cm} & {0.0\hspace{0.27cm}\,}\hspace{1.33cm} \\ 
${xz}^{1} \uparrow$ &  & -0.383 & {0.0\hspace{0.27cm}\,} &  & -0.363 +
0.056$i$ & -0.053\, --\, 0.100$i$ &  & -0.263 + 0.150$i$ & -0.142\, --\,
0.200$i$ \\ 
${yz}^{1} \uparrow$ &  & -0.383 & {0.0\hspace{0.27cm}\,} &  & -0.363\,
--\, 0.056$i$ & -0.100\, --\, 0.053$i$ &  & -0.263\, --\, 0.150$i$ & 
-0.200\, --\, 0.142$i$ \\ 
${xy}^{2} \uparrow$ &  & 0.454 & {0.0\hspace{0.27cm}\,} &  & 0.453\hspace{1.33cm} & {0.0\hspace{0.27cm}\,}\hspace{1.33cm} &  & 0.444\hspace{1.33cm} & {0.0\hspace{0.27cm}\,}\hspace{1.33cm} \\ 
${xz}^{2} \uparrow$ &  & 0.383 & {0.0\hspace{0.27cm}\,} &  & 0.363\, --\,
0.056$i$ & 0.053 + 0.100$i$ &  & 0.263\, --\, 0.150$i$ & 0.142 + 0.200$i$ \\ 
${yz}^{2} \uparrow$ &  & 0.383 & {0.0\hspace{0.27cm}\,} &  & 0.363 + 0.056$i$ & 0.100 + 0.053$i$ &  & 0.263 + 0.150$i$ & 0.200 + 0.142$i$ \\ 
${xy}^{1} \downarrow$ &  & {0.0\hspace{0.27cm}\,} & -0.454 &  & {0.0\hspace{0.27cm}\,}\hspace{1.33cm} & -0.453\hspace{1.33cm} &  & {0.0\hspace{0.27cm}\,}\hspace{1.33cm} & -0.444\hspace{1.33cm} \\ 
${xz}^{1} \downarrow$ &  & {0.0\hspace{0.27cm}\,} & -0.383 &  & 0.053\,
--\, 0.100$i$ & -0.363\, --\, 0.056$i$ &  & 0.142\, --\, 0.200$i$ & -0.263\,
--\, 0.150$i$ \\ 
${yz}^{1} \downarrow$ &  & {0.0\hspace{0.27cm}\,} & -0.383 &  & 0.100\,
--\, 0.053$i$ & -0.363 + 0.056$i$ &  & 0.200\, --\, 0.142$i$ & -0.263 + 0.150$i$ \\ 
${xy}^{2} \downarrow$ &  & {0.0\hspace{0.27cm}\,} & 0.454 &  & {0.0\hspace{0.27cm}\,}\hspace{1.33cm} & 0.453\hspace{1.33cm} &  & {0.0\hspace{0.27cm}\,}\hspace{1.33cm} & 0.444\hspace{1.33cm} \\ 
${xz}^{2} \downarrow$ &  & {0.0\hspace{0.27cm}\,} & 0.383 &  & -0.053 +
0.100$i$ & 0.363 + 0.056$i$ &  & -0.142 + 0.200$i$ & 0.263 + 0.150$i$ \\ 
${yz}^{2} \downarrow$ &  & {0.0\hspace{0.27cm}\,} & 0.383 &  & -0.100 +
0.053$i$ & 0.363\, --\, 0.056$i$ &  & -0.200 + 0.142$i$ & 0.263\, --\, 0.150$i$ \\ \hline
$j_{\text{eff}}=\frac{3}{2}$ weight &  & \multicolumn{2}{c}{0.6667} &  & 
\multicolumn{2}{c}{0.8320} &  & \multicolumn{2}{c}{0.9816} \\ 
$B_{1u}$ weight &  & \multicolumn{2}{c}{0.9932} &  & \multicolumn{2}{c}{
0.9567} &  & \multicolumn{2}{c}{0.7824} \\ \hline\hline
\end{tabular}
\end{center}
\end{table*}

The features shown in this Section, not unexpectedly, characterize
{\nio} as intermediate between the non-relativistic (pure
quasi-molecular orbital) and fully relativistic (pure RO)
cases.

 Moreover, these results show that, in the RO representation,
 the upper band states are not
pure $j_{\text{eff}}=1/2$ states but there is some significant mixing
of $j_{\text{eff}}=3/2$ states.
In fact, for the upper band states, the projections onto 
$j_{\mathrm{eff}}=1/2$ and 
$j_{\mathrm{eff}}=3/2$  are,
respectively, 0.64 and 0.21 
 with 2(0.64$^{2}+2\times 0.21^{2})=1$,
while in the non-relativistic case these projections are both equal to
$\sqrt{1/6}=0.41$.  Note that looking at the weights may be
misleading. Indeed this state appears to be 2$\times 0.64^{2}=82\%$
pure $j_{\text{eff}}=1/2$ state [Fig.~\ref{F.evolution}~(bottom)], but
its $projection$ on the $j_{\mathrm{eff}}=3/2$ state is only twice
smaller than in the non-relativistic case. In other words, the hopping
between the upper Kramers doublets, initially not considered in
Ref.~\onlinecite{Chaloupka2010}, is only reduced by about a factor of
two compared to the non-relativistic case. One but possibly not the
only consequence of this fact is that the contribution of the Kitaev
term in the analysis below may be overestimated, probably by as much
as a factor of two.

\subsection{Comparison with experiment: branching ratio}

An argument frequently used to justify the assumption of pure ROs in
{\nio} is that it is experimentally supported. However, the
experimental evidence is inconclusive. It is first assumed that the
electronic states are pure ROs and then it is shown that this
assumption does not contradict the experiment, yet the experiments,
upon a closer look, do not falsify the DFT picture, either. A typical and, by
far, the most often used quantity to discuss the nature of the states
in iridates is the branching ratio (BR) extracted from X-ray
absorption spectroscopy (XAS) experiments.  In XAS, essentially,
$\langle{\bf L}\cdot{\bf S}\rangle$
is measured. This expectation value is of course zero without
spin-orbit coupling. A detailed and very insightful analysis can be
found, for instance, in
Refs.~\onlinecite{Haskel,Chapon,Lovesey1,Lovesey2}. In particular, it
is shown that, for a related iridate, the main contribution to
$\langle{\bf L}\cdot{\bf S}\rangle$ $(1.4$ out of $2.1$) doesn't come
from the $t_{2g}$ orbitals, which define the $j_{\mathrm{eff}}=1/2$
states, but from the admixture of the $e_{g}$ orbitals. In our
calculations --shown below-- we observe the same behavior.

We apply our TB+SO model to calculate $\langle{\bf L}\cdot{\bf
  S}\rangle$ for {\nio} where ${\bf L}$ and
${\bf S}$ are, respectively, the \textit{total} orbital and spin
angular momenta of Ir $5d$ electrons.
  $\langle{\bf L}\cdot{\bf S}\rangle$ is related to
the experimentally accessible branching ratio as
\begin{equation}
\mbox{BR}=\frac{(2-r)}{(1+r)},\quad r=\frac{\langle{\bf L}\cdot{\bf S}\rangle }{n_{\text{h}}},
\end{equation}
with $n_{\text{h}}=5$ being the average number of $5d$ Ir
holes~\cite{Laan88,Thole88}. In recent XAS measurements~\cite{Clancy},
$\mbox{BR}=5.5-5.7$, translating to $\langle {\bf L}\cdot{\bf
  S}\rangle=-2.7\hbar^{2}$, was obtained for {\nio} and interpreted as
a sign of strong spin-orbit coupling.

When applying the TB+SO model that we constructed for {\nio} in
Section~\ref{sec:fiveA} the calculated
$\langle{\bf L}\cdot{\bf S}\rangle=-0.73\hbar^{2}$ (as compared to
$-1\hbar^{2}$ in the limit $\lambda_{\rm eff}=1$).  This value is several
times smaller than the experimental value. This is, however, not
unexpected given the significant contribution of the Ir~$e_{g}$ empty
states to $\langle{\bf L}\cdot{\bf S}\rangle$ ({\textit{cf.}}
Ref.~\onlinecite{Haskel}), which are not considered in the TB+SO model
discussed in the previous Section.  In order to make a meaningful
comparison with experiment, we extend our TB+SO model to include (in
the same spirit) also the Ir~$e_{g}$ states.  
$\langle{\bf L}\cdot{\bf S}\rangle$
 within such a model is $-1.91\hbar^{2}$.  This is
about 30\% less than the experimental value reported by
Clancy~{\textit{et~al.}}\cite{Clancy}. This result is indeed in good
agreement with experiment, given the large fluctuations in
experimental values. For instance, Ref.~\onlinecite{Clancy} reported
$\langle{\bf L}\cdot{\bf S}\rangle$ = $-3.1\hbar^{2}$ for
Sr$_{2}$IrO$_{4}$ while Ref.~\onlinecite{Haskel} reported $-2.1\hbar^{2}$
(about 30\% difference) for the same compound.  This example gives a
sense of possible fluctuations between results of different
experimental groups, and therefore our theoretical
$\langle{\bf L}\cdot{\bf S}\rangle$ value for {\nio} might be even
closer to the true result.

The main conclusion from these calculations is that with the TB+SO
model based on all {\it five} Ir~$5d$ orbitals we are able to
reasonably reproduce
 the
large experimentally measured $\langle{\bf L}\cdot{\bf S}\rangle$
value in {\nio}, which validates our approach. As our analysis shows,
the large $\langle{\bf L}\cdot{\bf S}\rangle$ does not necessarily
mean an ideal separation of $j_{\text{eff}}=\frac{3}{2}$ and
$j_{\text{eff}}=\frac{1}{2}$ RO states, but rather the effect
of  $e_g$ states also contributing in the process. Due to the
peculiar electron hopping hierarchy in {\nio}, QMOs
 might be a better basis.

 In
conclusion, the XAS experiments only tell us that the upper Kramers
doublet has a considerable contribution coming from
$j_{\mathrm{eff}}=1/2$, but not that it is a pure RO state.

\subsection{Comparison with experiment: RIXS}

Another experiment sometimes quoted as supporting the fully
relativistic $j_{\text{eff}}=\frac{1}{2}$ picture is resonant
inelastic x-ray scattering (RIXS)~\cite{RIXS}. In this experiment a
joint density of electronic states (JDOS) is probed, somewhat similar
to that in the infrared absorption but with different matrix
elements. The authors of Ref.~\onlinecite{RIXS} observed
several peaks in JDOS,
 of which the lowest peak at $\sim0.42$~eV was interpreted as
transitions across the Mott-Hubbard gap, consistent with a 30\%
smaller optical absorption threshold. The next two peaks are close to
each other at 0.72 and 0.83~eV and were ascribed to
transitions from the $j_{\text{eff}}=3/2$ quartet into the upper
$j_{\text{eff}}=1/2$ doublet. The splitting of 110~meV was ascribed to
the trigonal splitting.  Altogether, this interpretation
 suggests an SO coupling
$\lambda\sim\frac{2}{3}(\frac{0.72+0.83}{2}-\frac{0.42}{2})\,\text{eV}\approx0.39\,\text{eV}$,
a very reasonable number, if slightly too small.

This analysis, even though it looks reasonable on the first glance,
has serious shortcomings. First, the deduced trigonal splitting is
nearly twice as large as the actual trigonal splitting.  In fact,
the trigonal
splitting is decided by the electrostatic field of the ligands, and in addition
one-electron hoppings; both are very well accounted for
by the DFT calculations, which give $\Delta_T$ = 75~meV.  Second,
even a $\Delta_T= 110$~meV cannot produce well separated peaks in JDOS,
given that the Ir-Ir hopping is $t_{1\,{\rm O}}$ = 270~meV.  Third,
even if one completely neglects the Ir-Ir hoppings~\cite{RIXS}, in
order to extract $\lambda $ and $\Delta _{T}$ one has to diagonalize
the full Hamiltonian including both factors and then fit the resulting
eigenvalues to the observed peaks. After doing that, one gets $\lambda
=0.5$~eV and $\Delta _{T}=180$~meV. Although the previous
numbers are a rough estimate since they depend on the direction
of the Ir spins as well as on $U$ (here we considered $U=0$),
 the latter number is more than
twice the actual trigonal splitting.  
This argument shows that an interpretation of RIXS in
terms of infinitely narrow bands split by the trigonal field may not
be completely correct.

We find, on the other hand, that this experiment is  consistent with DFT band
structure. To demonstrate that, we have performed DFT calculations for the
magnetic zigzag phase.  We note that 
the results do not depend qualitatively on the
choice of the pattern and the magnetization direction.  
 In order to  account for the missing correlation
effects and adjust the direct gap to be consistent with infrared
measurements\cite{Comin12}, we  applied a
rigid shift of 200~meV between the occupied and empty bands (\textquotedblleft scissor
operator\textquotedblright ). 
 This exercise gives a JDOS which has a
broad feature, consisting of (i) a peak at 0.42 and a shoulder 0.48~eV
(compared to 0.42 eV in the experiment) corresponding to the transition
between the top QMOs and (ii) a peak at 0.77~eV and a shoulder at
0.81~eV corresponding to transitions from the lower QMOs. 
While the experiment finds
two peaks at 0.72 and 0.83~eV,  one should keep in mind that the
matrix elements,  omitted in our calculation, can easily suppress or
enhance a shoulder, making it disappear (at 0.48~eV) or become a
separate peak (at 0.81~eV). Therefore we conclude that the
agreement
between  experiment and our calculations, simplified as they are,  is reasonably good. 

\section{Magnetism}
\label{sec:six}

We proceed now with the discussion of the magnetic behavior of {\nio}.
Neutron diffraction experiments reported long-range 
antiferromagnetic order at low temperatures in a zigzag pattern~\cite{Radu}.
This ordering was confirmed by
 relativistic spin-polarized DFT calculations~\cite{we} where
we showed that it is the itinerancy of the system that
stabilizes the zigzag configuration. Such a pattern was also predicted
from the localized nnKH model~\cite{Chaloupka2010,Chaloupka2012}
 (Eq.~\ref{nnKHmodel}).
In the following we will provide {\it ab initio}-derived
estimates for the Kitaev and Heisenberg terms
and will show that in the
 physically reasonable parameter range this model unfortunately
fails to reproduce
the experimentally observed magnetic order.

\subsection{Nearest neighbor Kitaev-Heisenberg model}

One term neglected in the conventional Kitaev-Heisenberg model
treatment is  the single-site magnetocrystalline
anisotropy. Localized electrons with the spin $1/2$ do not have any
anisotropy, no matter how strong the spin-orbit coupling is. However, if
hopping is considered,  electrons can have
a preferred spin direction, which in the language of the nnKH
Hamiltonian would be reflected in a single-site term proportional, in
the lowest order, to 
$({\bf A}\cdot{\bf S})^2$ where ${\bf A}$ 
is a vector. Such terms are usually neglected when dealing with the
nnKH model.
Our calculations~\cite{we} without including $U$ show a magnetic
anisotropy as large as 3~meV per Ir (in order to address the
single-site anisotropy, we compared ferromagnetic calculations). This
energy should be compared to the total magnetic stabilization energy
 (i.e. the energy difference between magnetic and
non-magnetic solutions) 
of maximally 5~meV. When the DFT calculations are performed including
a $U=2$~eV, the magnetic anisotropy is as large as 8~meV out of a
total energy of 28~meV. This substantial anisotropy suggests that a
single site term should be added to the Kitaev-Heisenberg Hamiltonian,
probably resulting in a rather different phase diagram.

With all these caveats, it is still instructive to analyze where
{\nio} is to be found in the parametric space of the nnKH model.
We make  the following assumptions: (i) that the atomic orbitals
are fully localized and the appropriate basis is given by pure
$j_{\mathrm{eff}}=1/2$ orbitals; (ii) that the only hoppings relevant
for magnetic interactions are $pd$ hoppings, so that the only oxygen
assisted Ir-Ir hoppings are specific $t_{2g}-t_{2g}$ hoppings between
unlike orbitals, as outlined in Refs.~\onlinecite{Chaloupka2010,we},
and the $t_{2g}-e_{g}$ hoppings given in Ref.~\onlinecite{Chaloupka2012};
and (iii) that the only processes contributing to magnetic
interactions are those listed in
Ref.~\onlinecite{Chaloupka2012}. 

Indeed, the fact that  the experimentally
observed magnetic order is
zigzag suggests that either the Heisenberg terms are exceptionally
long ranged (the 3rd neighbor exchange is comparable to the 1st
one)~\cite{Radu,trebst}, or that the Kitaev term is strong and
antiferromagnetic~\cite{Chaloupka2012,chinese}.  The former suggestion
is seemingly in contradiction with the fact that the calculated 3rd
neighbor hoppings are substantially smaller than the 1st neighbor
ones. This makes it impossible to explain the large 3rd neighbor
exchange integral in terms of superexchange. However, there is a
possibility, suggested in Ref.~\onlinecite{we},  that the Ir electrons
are itinerant over individual hexagons, which makes magnetic
interactions naturally long ranged, and not directly related to the
hopping integrals.

The second suggestion,  which is the one we will focus on in what
follows, was proposed in Ref.~\onlinecite{Chaloupka2012}, namely that
of an antiferromagnetic Kitaev term. If strong enough, this could
explain the observed magnetic order. Below we consider the expressions
presented in Ref.~\onlinecite{Chaloupka2012} and substitute the
unknown variables with \textit{ab initio}-derived parameters.

Chaloupka \textit{et al.}~\cite{Chaloupka2012} discuss four relevant
processes contributing to the exchange interactions in {\nio}: (1)
Direct hopping $t_{1\sigma}$ between nearest neighbor Ir $t_{2g}$
orbitals contributing with a term $I_{1}=\left(
  \frac{2}{3}t_{1\sigma}\right) ^{2}/U$ to the Heisenberg term, where
$U$ is the Coulomb repulsion between $t_{2g}$ electrons.

(2) Interorbital nearest neighbor Ir $t_{2g}$-$e_{g}$ hopping via
intermediate oxygens $\tilde{t}_{1}$, with $\tilde{t}_{1}=t_{pd\sigma
}t_{pd\pi }/\Delta $, where $\Delta $ is the charge-transfer energy
(the difference between the O $p$ and Ir $d$ levels) contributing with
a term
$I_{2}=\frac{4}{9}\frac{\tilde{t}_{1}^{2}}{\tilde{U}}\,\frac{\tilde{J}_{\mathrm{H}}}{\tilde{U}}$
both to the Kitaev and Heisenberg terms, but with the opposite
signs. Here $\tilde{U}$ is the excitation energy associated with the
$t_{2g}$-$e_{g}$ hopping $i.e$. it also includes crystal field
splitting, $\tilde{U}$ = $U+10Dq$.  $\tilde{J}_{\mathrm{H}}$ is the
Hund's rule coupling between $t_{2g}$ and $e_{g}$ electrons.

(3) Oxygen-assisted hopping between two nearest neighbor Ir $t_{2g}$
orbitals $t_{1\,{\rm O}}$ contributing with a term
$I_{3}=\frac{8}{3}\frac{ t_{1\,{\rm
      O}}^{2}}{U}\,\frac{J_{\mathrm{H}}}{U}$ to the Kitaev term, where
$J_{\mathrm{H}}$ is the Hund's rule coupling between $t_{2g}$
electrons, and, we remind, $ t_{1\,{\rm O}} =t_{pd\pi }^{2}/\Delta $.

(4) Oxygen-$2p$ -- Iridium-$5d$ charge transfer contributing with a
term $I_{4}=\frac{8 t_{1\,{\rm O}}^{2}}{9}[\frac{1}{2\Delta
  +U_{p}-3J_{p}}+\frac{1}{3(2\Delta +U_{p}-J_{p})}+\frac{2}{3(2\Delta
  +U_{p}+2J_{p})}-\frac{1}{\Delta }],$ where $U_{p}$ and $J_{p}$ are,
respectively, the Hubbard repulsion and the Hund's rule parameter for
oxygen.~This expression was derived by G. Khaliullin~\cite{Khaliullin}
and is worth some additional discussion. The first three terms
correspond to processes where two holes of the same or of opposite
spins meet at an oxygen atom. Neglecting $J_{p},$ one gets simply
$\frac{8t_{1\,{\rm O}}^{2}}{9}\frac{1}{\Delta +U_{p}/2},$ which
reflects the fact that if the Ir atoms have opposite spins one can
create an intermediate state with two holes on the same oxygen
orbital, which lowers the total energy. The last term appears due to
ring exchange, with an intermediate state where two holes are located
on different oxygens. This process is only allowed when the ground
state is FM, and only if the ground state hole is in an $a_{1g}$ or
$j_{\rm eff}=1/2$ state, but not for pure $t_{2g}$ orbitals. However,
contrary to a common misconception, $J_{p}$ is  large, 
between 1.2 and 1.6~eV. We have estimated
$U_{p}$ and $J_{p}$, using the technique described in
Ref.~\onlinecite{petukhov}, and obtained $U_{p}= 2.7$ and $J_{p}= 1.6$~eV,
consistent with earlier DFT estimates~\cite{Mazin97}.  For non-relativistic orbitals
it is comparatively straightforward to account for the Hund's rule
coupling on O, but for relativistic orbitals it becomes more tedious.

If we expand  $I_4$  in both $U_{p}$ and $J_{p}$, then 
 $I_{4}\approx \frac{8 t_{1\,{\rm O}}^{2}}{9}
\frac{U_{p}-J_{p}}{2\Delta ^{2}}.$  This expression
shows that   $U_{p}$ alone contributes ferromagnetically to the
Heisenberg term and antiferromagnetically to the Kitaev term and may shift
the various phases in the nnKH model. Together with $J_{p}$ though, for the values
suggested above the effect of $U_{p}$ and
$J_{p}$ largely cancels and $I_4$  appears to be unimportant (note
though that if $J_{p}$ is entirely neglected, as in
Ref.~\onlinecite{Chaloupka2012},  this proposition becomes more questionable).

Summarizing the above terms into a single expression,
Eq.~(\ref{nnKHmodel}) can be written as:
\begin{equation}
H_{ij}^{(\gamma)}=\underbrace{(2I_{2}-I_{3}+2I_{4})}_{\displaystyle2K}S_{i}^{\gamma
}S_{j}^{\gamma }+\underbrace{(I_{1}-I_{2}-I_{4})}_{\displaystyle J}{\bf S}_{i}\cdot {\bf S}_{j}.  \label{eqKJ}
\end{equation}

This model has a zigzag magnetic ground state~\cite{Chaloupka2012} if
the Kitaev term is antiferromagnetic (AFM) and the Heisenberg term is
ferromagnetic (FM), with $K>0$, $J<0$ and
$-26\lesssim J/K \lesssim -0.3$.

In Table~\ref{tab:parameters} we provide the parameter values relevant
for {\nio}, as obtained from our DFT results. Note that the
$\tilde{t}_{1}$ parameter was assumed to be $2t_{1\,{\rm O}}$ in
Ref.~\onlinecite{Chaloupka2012}, while in the calculations (DFT
calculations are usually very reliable in this respect) 
$\tilde{t}_{1}/t_{1\,{\rm O}}$ is
1.4. However, using the ratio of 2 hardly changes any conclusions.

\begin{table}[ptb]
\begin{center}
\begin{tabular}{lll}
\hline\hline
parameter~ & value (eV)~ & meaning \\ \hline
$t_{1\sigma}$ & 0.03 & direct Ir-Ir hopping \\ 
 $t_{1\,{\rm O}}$  & 0.27 & O assisted Ir-Ir hopping \\ 
$\tilde{t}_1$ & 0.38 & Ir $t_{2g}-e_{g}$ hopping \\ 
$t_{pd\pi}$ & 0.57$^{*}$ & Ir-O $\pi$ hopping \\ 
$t_{pd\sigma}$&   1.6$^{*}$ & Ir-O $\sigma$ hopping\\
$\Delta$ & 2.4 & charge transfer energy\\ && between the O $p$ and Ir $d$ levels\\ 
$J_{\mathrm{H}}$ & 0.5 & Ir $t_{2g}$ Hund's rule coupling \\ 
$\tilde{J}_{\mathrm{H}}$ & 0.5 & Ir $t_{2g}-e_{g}$ Hund's rule coupling \\ 
\hline\hline
\end{tabular}
\end{center}
\caption{DFT-calculated values of transfer integrals and charge transfer
  energy for {\nio} and estimates of Hund's rule coupling
  strength as described in the text. The values marked with $*$ 
were obtained from $\tilde{t}_{1}$, $t_{1 O}$ and $\Delta$.}
\label{tab:parameters}
\end{table}

\begin{figure*}[tbh]
\includegraphics[width=0.8\textwidth]{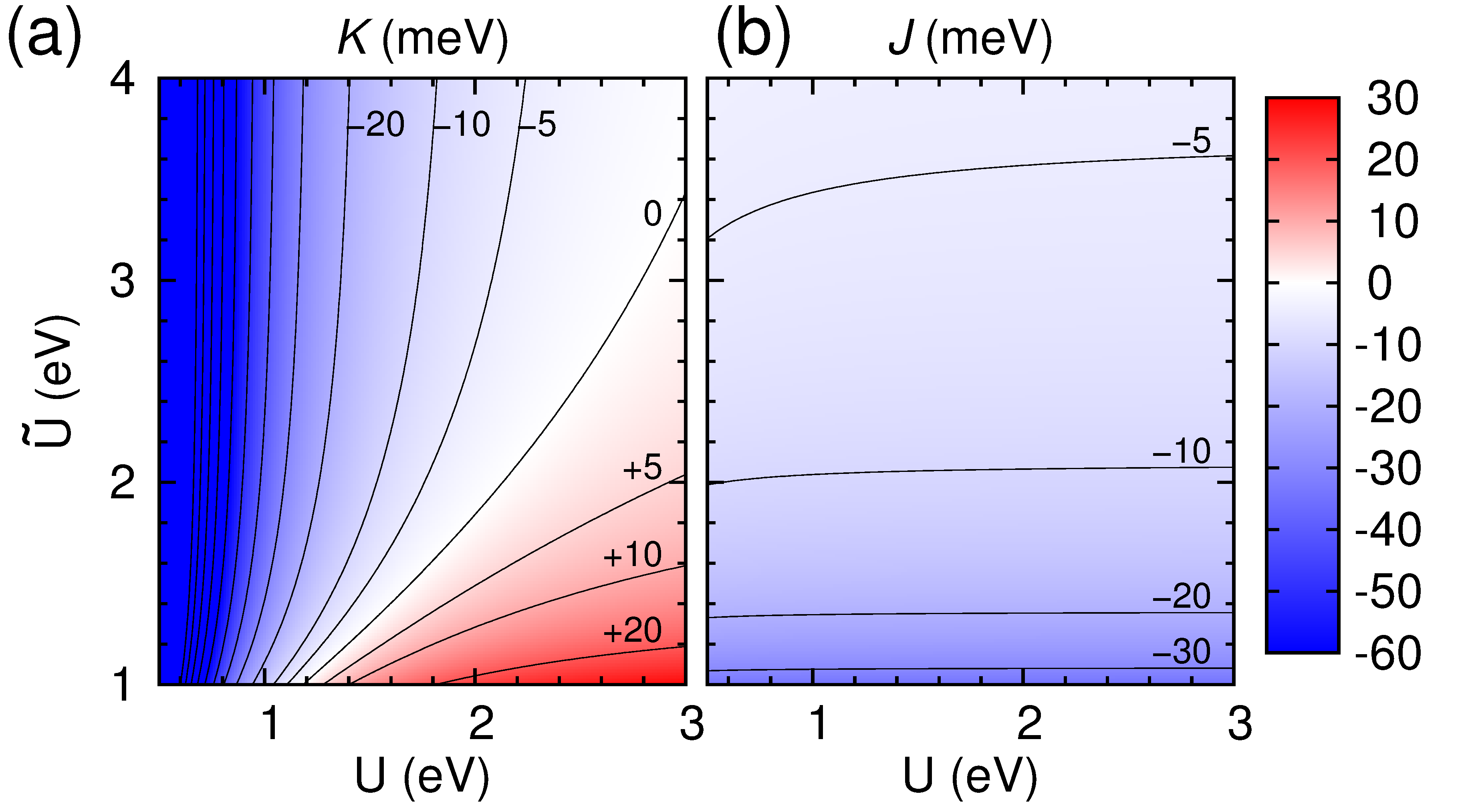}
\caption{(Color online) Variation of (a) Kitaev parameter $K$ and (b)
  Heisenberg exchange coupling $J$ with onsite Coulomb coupling
  strength $U$ and Ir $t_{2g}$-$e_{g}$ excitation energy
  $\tilde{U}$. Positive values refer to antiferromagnetic, negative to
  ferromagnetic values of $K$ and $J$. The other parameters entering
  the $K$ and $J$ are given in Table~\protect\ref{tab:parameters}. }
\label{KJ}
\end{figure*}

\begin{figure}[tbh]
  \caption{(Color online) Phase diagram of the Kitaev-Heisenberg model
    for {\nio} with parameters determined following
    Ref.~\onlinecite{Chaloupka2012}. The calculated exchange integrals
    are functions of the Mott-Hubbard gap $U$ and the cubic crystal
    field splitting $\tilde{U}$.  The contours mark isolines of the
    ratio $K/J$.}\label{all}
\includegraphics[width=\columnwidth]{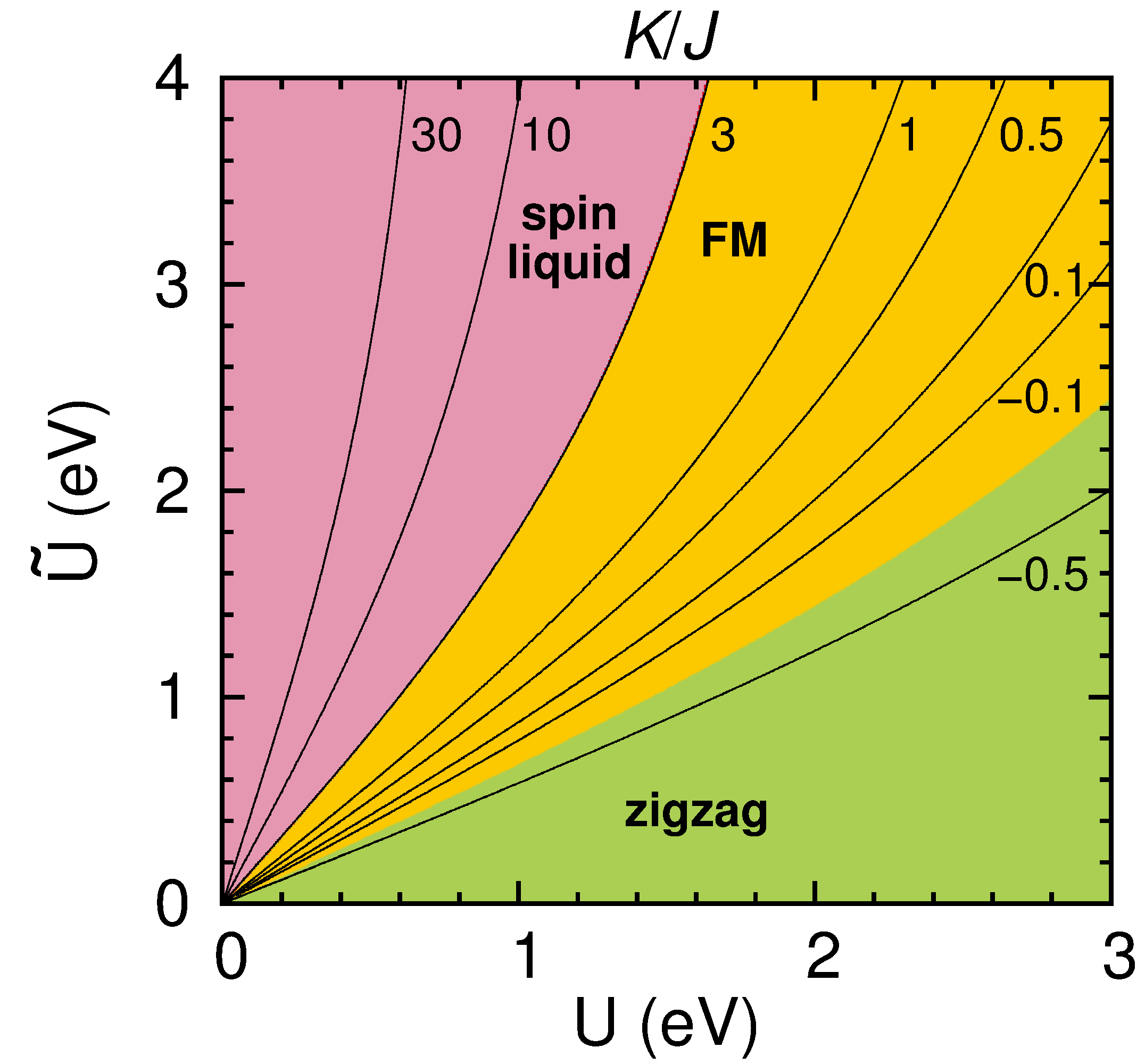}
\end{figure}

We present our results in Figures~\ref{KJ} and \ref{all}. In
Fig.~\ref{KJ} we show the calculated values of $K$ and $J$ as a
function of two variables: the $x$ axis is the Hubbard $U$ associated
with the upper Kramers doublet, and the $y$ axis is the energy
$\tilde{U}$, associated with exciting an individual electron from the
upper $t_{2g}$ to an average $e_{g}$ state. The Hubbard $U$ for $5d$
electrons is, generally speaking, 1.5 to 2~eV. However, in this case
it is additionally screened by the $e_{g}$ electrons, and also reduced
by hybridization (\textit{cf}. Na$_{x}$CoO$_{2}$~\cite{Liebsch} and Fe
pnictides~\cite{Georges}). Experimental estimates of the Hubbard $U$
defined as the energy cost for exciting electrons across the
insulating gap (which is the definition relevant to superexchange)
yield
0.3-0.5~eV~\cite{Comin12,RIXS}. 
Additionally, LDA+U calculations with $U\sim2 $~eV yield an excitation
gap of the same order. We conclude that the realistic range of this
parameter is 0.5--2~eV, with the smaller values more likely.

For the second parameter, $\tilde{U}$, DFT calculations give
$\sim2.5$~eV.  This should be considered as a lower bound since
 DFT tends
to slightly overestimate the orbital overlap and crystal fields, and
misses the effects of the $t_{2g}-e_{g}$ Hubbard interaction. One can
thus limit the physically admissible range in the region 2.5~eV
$\lesssim\tilde{U}\lesssim3$~eV.

In Figure~\ref{all} we show the phase diagram in the space of the two
parameters above. Several observations are in place: (1) While there
is a zigzag phase in this diagram, it is 
very far removed from the
range of the parameters that can be called physical, 0.5~eV $\lesssim
U\lesssim2$~eV, 2~eV$\lesssim$ $\tilde{U}\lesssim3$~eV (even though in
the above estimate we have liberally stretched the admissible range in
favor of a zigzag phase).  In fact, the zigzag regime appears only when
$\tilde{U}<0.6U$, i.e. when the Hubbard gap is larger than the
$e_{g}-t_{2g}$ splitting, a rather unlikely proposition. (2) In the
physical range of parameters, the ground state is either ferromagnetic
or the spin liquid phase. It is rather curious that the very narrow
slivers of the phase space in the $J,K$
coordinates~\cite{Chaloupka2012} are transformed into a very large
range in the $U,\tilde{U}$ space.

It is also worth mentioning that in order to explain the experimental
data of Refs.~\onlinecite{Radu,trebst} one needs not only the ground
state to be zigzag, but also that $K$ be several times larger than
$|J|$; Chaloupka \textit{et al.}~\cite{Chaloupka2012} used $K=10.44$
and $J=-4.01$~meV. This solution cannot be obtained for a given set of $U$ and $\tilde{U}$ (see Fig.~\ref{KJ}). Moreover, a closer look at the expressions in their
work reveals that $K+J=I_{1}-I_{3}/2$, which does not depend on
$\tilde{U}$ and is always negative.  Thus the two equalities above
cannot be satisfied simultaneously for any choice of parameters, be
they physical or not. Moreover, the values of $J$ and $K$ used in
Ref.~\onlinecite{Chaloupka2012} can only be obtained if
$\tilde{U}<0.2$~eV, which is clearly an impossible  regime.

\subsection{Long-range exchange}

As mentioned above, an alternative interpretation of the experimental
results, given in Refs.~\onlinecite{Radu,trebst}, is in terms of
sizable 2nd and 3rd neighbor exchange constants, comparable to the
nearest neighbor exchange. In this picture the Kitaev term may or may
not play a role, but this role is not decisive in establishing the
observed magnetic order. Given that the calculated hopping amplitudes
(Table I) are clearly dominated by the nearest neighbor terms,
standard superexchange cannot explain such long range
interactions.

However, it is important to remember that in the opposite, itinerant
limit every electron is fully delocalized over a hexagon and, as such,
is equally sensitive to the mean field magnetization pattern on the
1st, 2nd or 3rd nearest neighbors. As discussed in our earlier
work~\cite{we}, the zigzag order, as compared to the stripy one,
results in a sizable pseudogap at the Fermi level even without a
Hubbard $U$. This creates an energy gain that cannot be cast in a form
of nearest neighbor interaction, as it depends on the magnetization
pattern over an entire hexagon.

We are far from stating that the superexchange Hamiltonian outlined in
Ref.~\onlinecite{Chaloupka2012} is irrelevant and an itinerant
description will give the final answer to all questions regarding the
magnetism in {\nio}. However, relying solely on the localized picture
and, correspondingly, on the nnKH model, is, apparently, inadequate.

\section{Conclusions}

In summary, we have performed an extensive investigation
of the electronic properties of {\nio} in the framework
of non-relativistic and relativistic density functional theory
calculations and derived by means of the Wannier function
projector method, the corresponding microscopic parameters.
We resolved the following open questions: 
(1) By considering
various idealized crystal structures for {\nio} we could
disentangle the effect of each of the structural
distortions present
in this system and concluded that it is the joint effect
of these distortions that constructively enhances the  
 intrahexagon effective hopping parameters and suppresses
the interhexagon ones favoring the formation
of quasi-molecular orbitals. (2) We modelled the relativistic
DFT results in terms of a tight-binding model including the spin-orbit
coupling term and analyzed the electronic properties of {\nio} in terms
of two complementary descriptions, the (itinerant)
quasi-molecular basis and the (localized) relativistic $j_{\rm eff}$
basis. We observed that the behavior of {\nio} lies in between
the fully itinerant and the fully localized description and that
a quasi-molecular orbital description keeps its character even 
at large values of the spin-orbit coupling strength. (3) We
showed that XAS and RIXS observations can be well understood
within an itinerant description 
of {\nio} in contrast to other iridates like Sr$_3$CuIrO$_6$ where
localization is imposed by the crystallographic arrangement
of the IrO$_6$ octahedra~\cite{Liu2012}. (4) Finally, we provided {\it ab initio}-derived estimates
for the parameters appearing in the Kitaev and Heisenberg
terms in {\nio} and found that the recently proposed nnKH
model (see Section~\ref{sec:six}), even though  it is a very
interesting model {\it per se},  is unfortunately
 not realistic for {\nio}.  In conclusion, 
in order to obtain a full understanding of the
behavior of {\nio}  all three features; spin-orbit, Coulomb
correlations and delocalization of valence electrons over Ir$_6$
hexagons  are essential. 

 H.O.J., D.Kh. and R.V. acknowledge support by the Deutsche
Forschungsgemeinschaft through grants SFB/TR 49 and FOR 1346
(H.O.J. and R.V.) and SFB 608 and FOR 1346
(D.Kh.).
\label{sec:seven}

\section{Appendix. Idealized {N\lowercase{a}$_{2}$I\lowercase{r}O$_{3}$} crystal structures as used in WIEN2\lowercase{k}}

\subsection{Experimental structure from Ref.~\onlinecite{Radu}}
\begin{center}
\begin{tabular}{cp{1cm}cp{1.0cm}cp{0.6cm}c}
\hline\hline
\multicolumn{7}{c}{Space group $C2/m$ (No.~12)} \\ 
\multicolumn{7}{c}{$a=5.4269$~\AA ,\; $b=6.4104$~\AA ,\; $c=9.3949$~\AA , $\gamma=124.12^\circ$} \\ \hline\hline
Atom &  & $x$ &  & $y$ &  & $z$ \\ \hline
Na1 &  & 0.0 &  & 0.0 &  & 0.5 \\ 
Na2 &  & 0.5 &  & 0.5 &  & 0.0 \\ 
Na3 &  & 0.5 &  & 0.5 &  & 0.3400 \\ 
Ir &  & 0.0 &  & 0.0 &  & 0.1670 \\ 
O1 &  & 0.4590 &  & 0.2110 &  & 0.1780 \\ 
O2 &  & 0.0070 &  & 0.7960 &  & 0.0 \\ \hline\hline
\end{tabular}
\end{center}

\subsection{Idealized structure $S_1$}
\begin{center}
\begin{tabular}{cp{1cm}cp{1.0cm}cp{0.6cm}c}
\hline\hline
\multicolumn{7}{c}{Space group $C2/m$ (No.~12)} \\ 
\multicolumn{7}{c}{$a=5.4501$~\AA ,\; $b=6.4411$~\AA ,\; $c=9.4399$~\AA , $\gamma=125.26^\circ$} \\ \hline\hline
Atom &  & $x$ &  & $y$ &  & $z$ \\ \hline
Na1 &  & 0.0 &  & 0.0 &  & 0.5 \\ 
Na2 &  & 0.5 &  & 0.5 &  & 0.0 \\ 
Na3 &  & 0.5 &  & 0.5 &  & 0.3333 \\ 
Ir &  & 0.0 &  & 0.0 &  & 0.1667 \\ 
O1 &  & 0.4646 &  & 0.2097 &  & 0.1785 \\ 
O2 &  & 0.0000 &  & 0.7903 &  & 0.0 \\ \hline\hline
\end{tabular}
\end{center}

\subsection{Idealized structure $S_2$}
\begin{center}
\begin{tabular}{cp{1cm}cp{1.0cm}cp{0.6cm}c}
\hline\hline
\multicolumn{7}{c}{Space group $C2/m$ (No.~12)} \\ 
\multicolumn{7}{c}{$a=5.4501$~\AA ,\; $b=6.4190$~\AA ,\; $c=9.4399$~\AA , $\gamma=124.47^\circ$} \\ \hline\hline
Atom &  & $x$ &  & $y$ &  & $z$ \\ \hline
Na1 &  & 0.0 &  & 0.0 &  & 0.5 \\ 
Na2 &  & 0.5 &  & 0.5 &  & 0.0 \\ 
Na3 &  & 0.5 &  & 0.5 &  & 0.3333 \\ 
Ir &  & 0.0 &  & 0.0 &  & 0.1667 \\ 
O1 &  & 0.4606 &  & 0.1909 &  & 0.1667 \\ 
O2 &  & 0.0000 &  & 0.8091 &  & 0.0 \\ \hline\hline
\end{tabular}
\end{center}

\subsection{Idealized structure $S_3$}
\begin{center}
\begin{tabular}{cp{1cm}cp{1.0cm}cp{0.6cm}c}
\hline\hline
\multicolumn{7}{c}{Space group $C2/m$ (No.~12)} \\ 
\multicolumn{7}{c}{$a=5.0658$~\AA ,\; $b=5.9869$~\AA ,\; $c=8.7743$~\AA , $\gamma=125.26^\circ$} \\ \hline\hline
Atom &  & $x$ &  & $y$ &  & $z$ \\ \hline
Na1 &  & 0.0 &  & 0.0 &  & 0.5 \\ 
Na2 &  & 0.5 &  & 0.5 &  & 0.0 \\ 
Na3 &  & 0.5 &  & 0.5 &  & 0.3333 \\ 
Ir &  & 0.0 &  & 0.0 &  & 0.1667 \\ 
O1 &  & 0.5000 &  & 0.2443 &  & 0.1667 \\ 
O2 &  & 0.0000 &  & 0.7557 &  & 0.0 \\ \hline\hline
\end{tabular}
\end{center}

Note that due to the necessity of using a monoclinic angle $\gamma$ in
WIEN2k, the Ir honeycomb layers in the {\nio} unit cells presented
above are parallel to the $ac$ plane. Accordingly, within this
convention the vector of the Bloch factors in Eqs.~(\ref{E.QMOproj})
and (\ref{E.SOtrans}) is given by
\begin{equation}
T_{M=1\ldots6}({\bf k})=
(1, e^{-ik_{x}\tilde{a}}, e^{ik_{z}\tilde{c}},
e^{i(k_{z}\tilde{c}-k_{x}\tilde{a})}, e^{ik_{z}\tilde{c}}, e^{-ik_{x}\tilde{a}}),
\end{equation}
where $\tilde{a}$ and $\tilde{c}$ are the lengths of the two primitive
lattice vectors lying in the $ac$ plane.  Here, one explicitly
accounts for the choice of WIEN2k of the actual positions of the two
Ir atoms in the primitive unit cell, which are, {\eg},
(-0.167,0,0.167) and (-0.833,0,0.833) in the experimental {\nio}
structure.


\begin{thebibliography}{99}
\bibitem{comment_split} The spin-orbit splitting  is given
by the difference of the eigenvalues $\lambda /2$
and $-\lambda$ in the $j_{\rm eff}$ space.
\bibitem{axes} The description of the orbital character in the present
  work is done according to the local cubic coordinate system if not
  explicitly specified otherwise.
\bibitem{note1} In Fig.~\ref{F.Wannier} we visualize real-valued
  combinations of degenerate $E_{2u}$ or $E_{1g}$ QMOs.
\bibitem{comment_basis} Note that these states are eigenstates of
  Eq.~(\ref{matrix_SO}) after diagonalizing the matrix
  Eq.~(\ref{matrix_SO}) and performing an appropriate rotation.
\end{thebibliography}
\end{document}